\documentclass[%
 reprint,
superscriptaddress,
 amsmath,amssymb,
 aps,
prl
floatfix,
]{revtex4-1}

\usepackage{graphicx}% Include figure files
\usepackage{dcolumn}% Align table columns on decimal point
\usepackage{bm}% bold math
\usepackage{subfigure}
\usepackage{hyperref}% add hypertext capabilities
\usepackage{wrapfig}
\usepackage{placeins}
\usepackage{natbib}
\usepackage{array,booktabs,longtable,tabularx,tabulary}
\usepackage{amsmath}
\usepackage{tikz}
\usetikzlibrary{positioning,shapes}
\begin{document}

%\preprint{APS/123-QED}

\title{The dipolar spin glass transition in three dimensions}

\author{Tushar Kanti Bose}
%\email{tkb.tkbose@gmail.com}
\affiliation{School of Physical Sciences, Indian Association for the Cultivation of Science, 2A \& 2B Raja SC Mullick Road, Jadavpur, Kolkata-700032, India.}
\author{Roderich Moessner}
%\email{moessner@pks.mpg.de}
\affiliation{Max Planck Institute for the Physics of Complex Systems, N\"{o}thnitzer Strasse 38, 01187 Dresden, Germany}
\author{Arnab Sen}
%\email{tpars@iacs.res.in}
\affiliation{School of Physical Sciences, Indian Association for the Cultivation of Science, 2A \& 2B Raja SC Mullick Road, Jadavpur, Kolkata-700032, India.}%Lines break automatically or can be forced with \\

%\affiliation{
%\affiliation{
%\affiliation{%


\date{\today}% It is always \today, today,

\begin{abstract}
Dilute dipolar Ising magnets remain a notoriously hard problem to tackle both 
analytically and numerically because of long-ranged interactions 
between spins as well as rare region effects. We study a new type of 
anisotropic dilute dipolar Ising system in three dimensions 
[Phys. Rev. Lett. {\bf 114}, 247207 (2015)] 
that arises as an effective description of  
randomly diluted classical spin ice, a prototypical spin liquid in the 
disorder-free limit, with a small fraction $x$ of non-magnetic impurities. 
Metropolis algorithm within a parallel thermal tempering scheme 
fails to achieve equilibration for this problem already for small
system sizes. Motivated by previous work [Phys. Rev. X {\bf 4}, 041016 (2014)]
on uniaxial random dipoles, 
we present an improved cluster Monte Carlo algorithm that is 
tailor-made for removing the equilibration bottlenecks 
created by clusters of {\it effectively frozen} spins.  
By performing large-scale simulations down to $x=1/128$ and using 
finite size scaling, we show the existence of a finite-temperature 
spin glass transition and give strong evidence that the universality of 
the critical point is independent of $x$ when it is small. In this 
$x \ll 1$ limit, we also provide a first estimate of both the thermal 
exponent, $\nu=1.27(8)$, and 
the anomalous exponent, $\eta=0.228(35)$. 
\end{abstract}

\pacs{Valid PACS appear here}% PACS, the Physics and Astronomy
\maketitle

%\tableofcontents

%%%%Introduction section
{\underline{{\it Introduction:}}}
The term ``spin glass'' was originally coined to describe dilute 
magnetic alloys like AuFe~\cite{CannellaM72} composed of non-magnetic 
metals (like Au) weakly 
diluted with magnetic impurities (like Fe) where the impurity spins 
interact with an RKKY exchange~\cite{RKKY}.  
Since then, glassy behaviour of spins has 
been realized in a variety of magnetic systems~\cite{BinderY86} 
and the general wisdom is that 
both frustration and disorder are necessary ingredients for glassiness. 
Most theoretical studies have focused on Edwards-Anderson type 
models~\cite{EdwardsA75} where the spin interactions are short-ranged 
and random in sign. Extensive numerical simulations have now established 
the presence of a finite-temperature spin glass transition for Ising spins in 
three dimensions and its associated 
critical exponents have been accurately 
computed ~\cite{EAnumerics1, EAnumerics2}.  
Such Ising systems are, however, experimentally rare where it is 
much more common to have Heisenberg spins~\cite{BinderY86,Mydoshbook}. 
With these isotropic degrees of freedom, the nature of the transition is 
still controversial in the corresponding Edwards-Anderson 
model~\cite{vectormodelnumerics1, 
vectormodelnumerics2, vectormodelnumerics3, vectormodelnumerics4}. 

Disordered dipolar Ising magnets such as LiHo$_x$Y$_{1-x}$F$_4$ 
provide another class of candidate systems~\cite{review_lithium}, 
qualitatively distinct from their short-ranged counterparts, 
where the twin ingredients of frustration (due to the nature of the 
magnetostatic dipole-dipole interaction) and disorder (in the 
spatial arrangement of the magnetic ions like Ho$^{3+}$ when randomly 
substituted by non-magnetic Y$^{3+}$ ions) are both present. 
Experiments have found a spin glass phase for $x<x_c$~\cite{ReichEYRAB90} 
where $x_c \approx 0.25$ at low temperature. 
An analysis of the ac susceptibility shows that a spin glass phase 
may exist even at extreme dilutions of $x=0.045$~\cite{QuilliamMMK08} 
suggesting that a finite-temperature 
spin glass phase may extend all the way from $x_c$ 
down to $x \rightarrow 0^+$.

However, unlike their short-ranged counterparts, 
the nature of the spin glass ordering in dilute dipolar Ising
systems has been a long-standing open issue as conventional analytical 
and numerical techniques suffer different problems, 
particularly in the high dilution limit.
Mean-field theory suggests that the spin glass order is maintained even in the 
high-dilution limit, with the critical temperature being linear in the 
concentration of the spins~\cite{StephenA81, XuBNR91}. 
However, at high dilution, spatial inhomogeneities are large and could 
easily modify the mean-field theory predictions. 
In numerical simulations, long equilibration 
times severely limit the studied system sizes and concentrations. 
Even in experiments, equilibration is difficult to achieve 
due to ultraslow dynamics above the transition temperature (which may be 
around $10^7$ slower than in short-ranged spin 
glass materials~\cite{BiltmoH12}). 
Due to these difficulties, even the existence of the spin glass transition in 
such magnets has been a matter of 
long-standing debate~\cite{DSG1,DSG2,DSG3,DSG4}. 
Recent large-scale numerical simulations have shown a 
spin glass phase down to experimentally relevant 
low concentrations~\cite{Gingras, AndresenKOS14}. 
The universality class of the highly dilute dipolar Ising magnet 
in three dimensions, though, is still unknown.

In this work, we study a different, experimentally motivated, example of 
an {\it emergent} dilute anisotropic dipolar system that arises 
on {\it weakly} diluting dipolar spin ice materials on the three dimensional 
pyrochlore lattice of corner-sharing tetrahedra
with non-magnetic impurities, e.g., 
Dy$_{2-x}$Y$_x$Ti$_2$O$_7$/Ho$_{2-x}$Y$_x$Ti$_2$O$_7$,
where the magnetic Dy$^{3+}$/Ho$^{3+}$ ions 
are replaced randomly 
by non-magnetic Y$^{3+}$ ions~\cite{KeFULDCMS07, LinKTSMG14}. 
The disorder-free problem is known to exhibit a 
topological Coulomb phase~\cite{CastelnovoMS12} 
characterised by several non-trivial features like a 
pinch-point motif in the spin structure 
factor~\cite{IsakovGMS04, Henley05, SenMS13}, 
large residual entropy of the spins at low temperature~\cite{RamirezHCSS99} 
and emergent magnetic monopoles~\cite{CastelnovoMS08}. 
%In the highly dilute limit ($x \rightarrow 1^{-}$), the remaining Ising spins, whose orientation are
%neither random nor collinear but being picked from the local axes of the occupied sites (thus respecting 
%the cubic symmetry of the original pyrochlore lattice), interact via magnetostatic dipole-dipole interactions.
In the weak dilution limit ($x \ll 1$), it was shown in 
Ref.~\onlinecite{SenM15} that the dense but disordered network of 
Ising spins can be mapped to a 
dilute network of {\it emergent} Ising spins (dubbed ghost spins) 
that reside on the sites of the missing spins, have the same 
local Ising easy axes as the corresponding missing spins (the cubic symmetry 
of the pyrochlore lattice does not allow collinear Ising axes; rather, each 
spin points along the line joining the centres of two connected tetrahedra), 
and are again coupled by a magnetostatic dipole-dipole interaction
but with a renormalised coupling constant. We note that, statistically, 
this model retains the full cubic symmetry of the pyrochlore lattice.
%The universality class of a possible spin glass 
%transition is thus the same in both the weakly and highly disordered limits 
%of diluted spin ice, while the corresponding transition temperatures 
%are related by $T_c(x) = T_x/ \left( 1-\frac{3T_x}{\sqrt{2\pi}}\right)$ 
%where $T_c(x)$ ($T_x$) is the transition temperature in the weakly (highly) 
%dilute limit~\cite{SenM15}.  

We perform large-scale numerical simulations of this effective 
dilute dipolar magnetic system using an improved version of a 
cluster algorithm used in Ref.~\onlinecite{AndresenKOS14} (the 
basic idea was also introduced previously in 
Refs.~\onlinecite{JanzenHE08, AndresenJK11}) 
to establish the presence of a finite-temperature spin glass transition.
Our algorithm differs from Ref.~\onlinecite{AndresenKOS14} both in the 
definition of a cluster and in the relative importance associated to different 
clusters during a Monte Carlo step. 
We comment on the relation of our cluster construction to dynamically 
frozen spin clusters. We also discuss the manner in which the 
efficiency of the algorithm may be controlled by tuning the cluster 
construction parameters since it is not rejection-free 
(unlike Swendsen-Wang~\cite{SW} and Wolff~\cite{Wolff} algorithms for 
unfrustrated spin models). 

Using our cluster algorithm, 
we have reached total number of spins, $N$, that are roughly 
twice as large compared to the previous large-scale 
simulations of related uniaxial Ising systems~\cite{AndresenKOS14}
(note that 
the CPU time increases quadratically with $N$ for every Monte Carlo sweep 
of the system owing to long-ranged interactions 
between the spins). 
When $x \ll 1$, this problem provides a 
{\it different} lattice realization of (presumably) 
the {\it same} universal physics of the 
spin glass transition as uniaxial Ising spins interacting via a dipolar 
coupling in the dilute limit.
Using finite-size scaling, we show that $T_x \propto x$ and 
the universality class of the transition is independent of $x$ 
when it is small. The study of this model enables us to provide an 
estimate of both the thermal exponent $\nu$ and the anomalous 
exponent $\eta$ at small $x$ unlike the unixial dipolar model 
studied earlier in Refs.~\onlinecite{Gingras,AndresenKOS14} (where only $\nu$ 
could be reliably estimated).

%%%%%%%%%%%%%%%%%%% The Model

{\underline{{\it The Model:}}} In dipolar spin ice, 
the Ising spins on the pyrochlore lattice have 
local easy axis directions, $\hat{e}_i$, that are defined by the line 
joining the centers of the pair of tetrahedra which share them. 
The simplest appropriate interaction Hamiltonian of Ising spins with 
moments $\vec\mu_{i,j}$ of size $\mu$, separated by $r_{ij}$, 
contains short-range exchange interactions in addition to the 
usual magnetic dipolar term, $D{\cal D}_{ij}$, of strength 
$D$,with
%%%%%%%%%%%%%%%%%%%%%%%%%%%%%%%
\begin{equation}
{\cal D}_{ij}=
%\left( D + \frac{3T}{\sqrt{2} \pi} \right) 
\frac1{\mu^2}\left( \frac{a}{r_{ij}} \right)^3 \left( \vec\mu_i\cdot\vec\mu_j - 3 ( \vec\mu_i\cdot\hat{r}_{ij})( \vec\mu_j\cdot\hat{r}_{ij})\right) 
\label{eq:effint1}
\end{equation} 
%%%%%%%%%%%%%%%%%%%%%%%%%%%%%%%
where $a$ is the nearest neighbor distance on the pyrochlore 
lattice~\cite{CastelnovoMS12}.
%and $a_d=a\sqrt{3/2}$ the separation between the centers of the tetrahedra, 
%which define a diamond lattice 

Following Ref.~\onlinecite{SenM15}, a weakly 
diluted system of spins can be mapped to a highly diluted system of 
emergent ghost spins. The pairwise interaction between the ghost spins, 
$\tilde{\cal H}_{ij}$ has the standard dipolar form 
$\tilde{{\cal H}}_{ij}=\tilde{D}{\cal D}_{ij}$ where 
%%%%%%%%%%%%%%%%%%%%%%%%%%%%%%%
%\begin{equation}
%\tilde{{\cal H}}_{ij}=\tilde{D}{\cal D}_{ij}
%\label{eq:effint2}
%\end{equation}  
%%%%%%%%%%%%%%%%%%%%%%%%%%%%%%%
\(\tilde{D}\) is the effective dipolar coupling constant between the 
ghost spins which has an entropic contribution coming from the 
fluctuations of the spins in the bulk~\cite{SenM15} on top of the simple 
magnetostatic coupling constant $D$:
%%%%%%%%%%%%%%%%%%%%%%%%%%%%%%%
\begin{equation}
\tilde{D}=D + \frac{3T}{\sqrt{2} \pi}
\label{eq:effD}
\end{equation}
%%%%%%%%%%%%%%%%%%%%%%%%%%%%%%%

Henceforth, we will consider the dipolar coupling constant to be set to 
$D$=1.41 K (as in Ho$_{2}$Ti$_{2}$O$_{7}$ and Dy$_{2}$Ti$_{2}$O$_{7}$
~\cite{Sc294_2001}). The renormalization of $D$ to $\tilde{D}$ simply 
renormalizes the transition temperature to be 
$T_{c}(x)=T_{x}/(1-\frac{3T_{x}}{\sqrt 2 \pi})$ where $T_x$ is the transition 
temperature with the coupling set to be $D$. Here $x$ denotes the density 
of the ghost spins (which is assumed to be small). 
Thus, the Hamiltonian $\mathcal{H}$ that is studied numerically in 
this work has the form: 
%%%%%%%%%%%%%%%%%%%%%%%%%%%%%%%
\begin{eqnarray}
{\cal H}&=&
%\left( D + \frac{3T}{\sqrt{2} \pi} \right) 
D\sum_{i,j(i>j)}\left[\left( \frac{a}{r_{ij}} \right)^3 \left( \hat{e}_i\cdot\hat{e}_j - 3 ( \hat{e}_i\cdot\hat{r}_{ij})( \hat{e}_j\cdot\hat{r}_{ij})\right)\right]S_i S_j \nonumber \\
&=& \sum_{i,j(i>j)} J_{ij}S_iS_j
\label{eq:Ising}
\end{eqnarray} 
%%%%%%%%%%%%%%%%%%%%%%%%%%%%%%%
where $\mu_i = \mu S_i \hat{e}_i$ with $S_i=\pm 1$. The long-ranged nature of 
the dipolar interactions is treated using the 
Ewald summation technique~\cite{Ewald} without a demagnetization factor.

%%%%%%%%%%%%%%%%%%%%%%%%%%%%%%%%%%%%%%%%%%%%%%%%%%%%%
\begin{figure*}[t]
\centering
\subfigure[\label{fig1a} $T=0.072 \approx 2T_x$]{\includegraphics[scale=0.25]{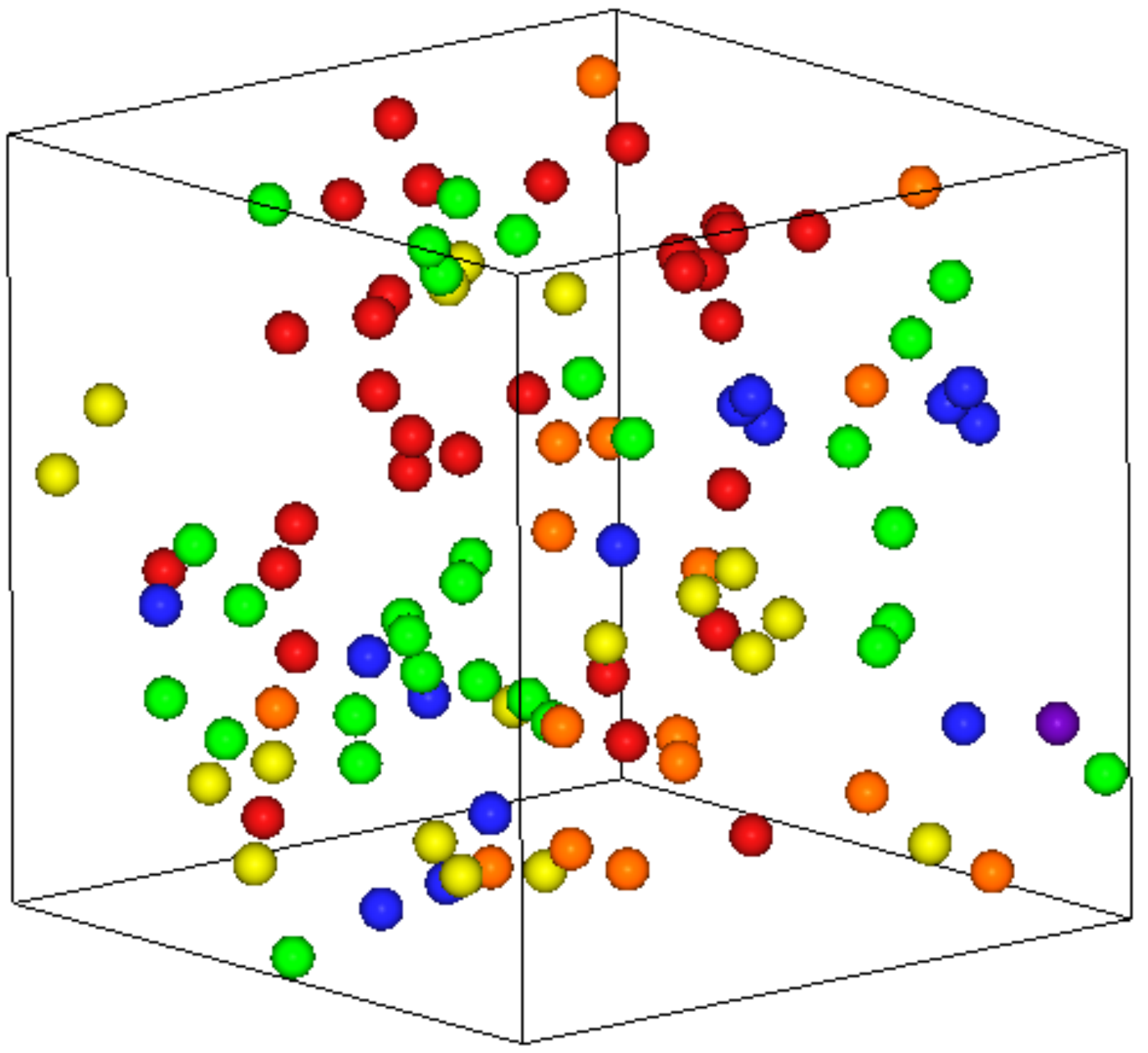}}
\subfigure[\label{fig1b} $T=0.047 \approx 1.3 T_x$]{\includegraphics[scale=0.25]{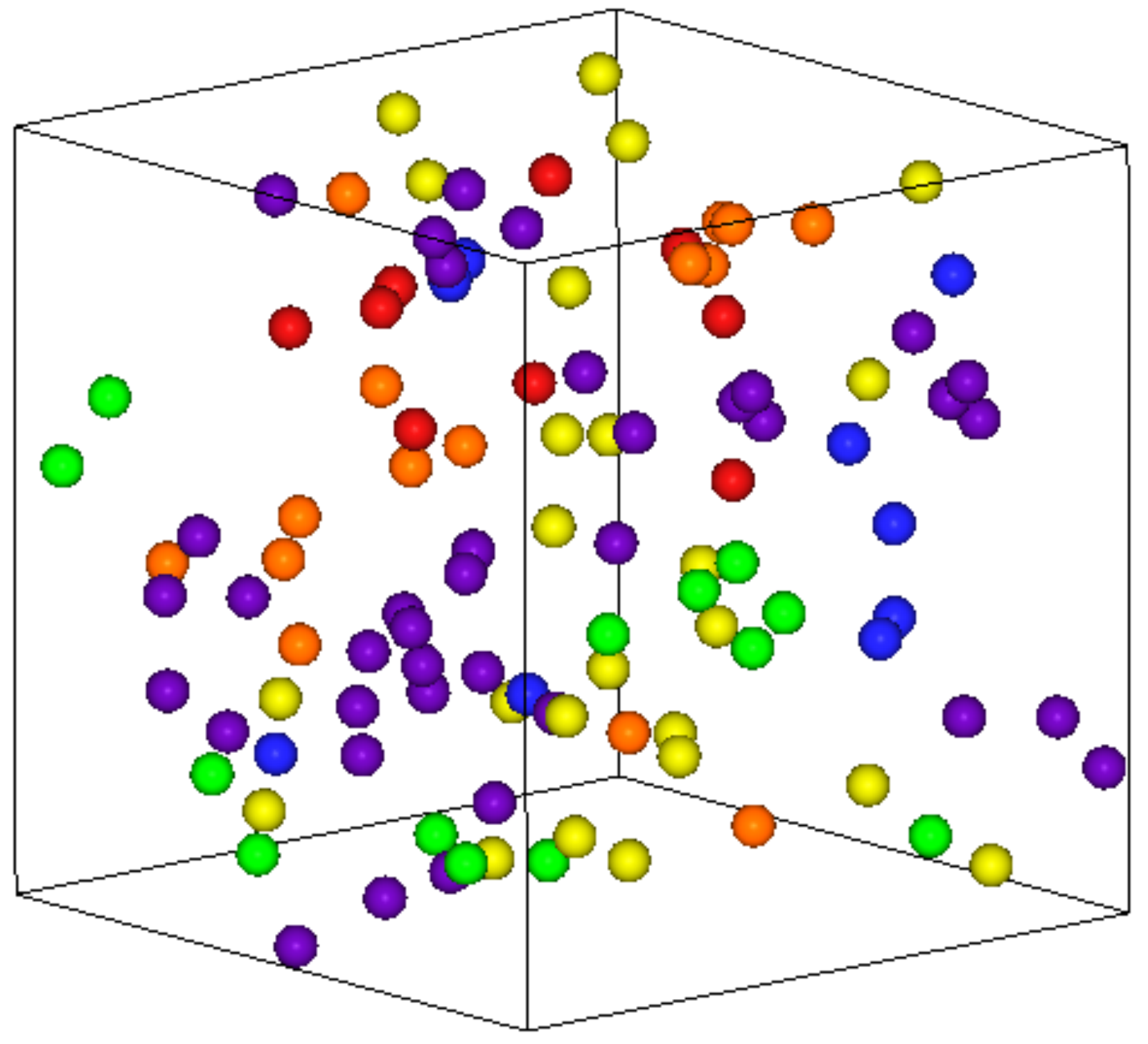}}
\subfigure[\label{fig1c} $C_2$]{\includegraphics[scale=0.25]{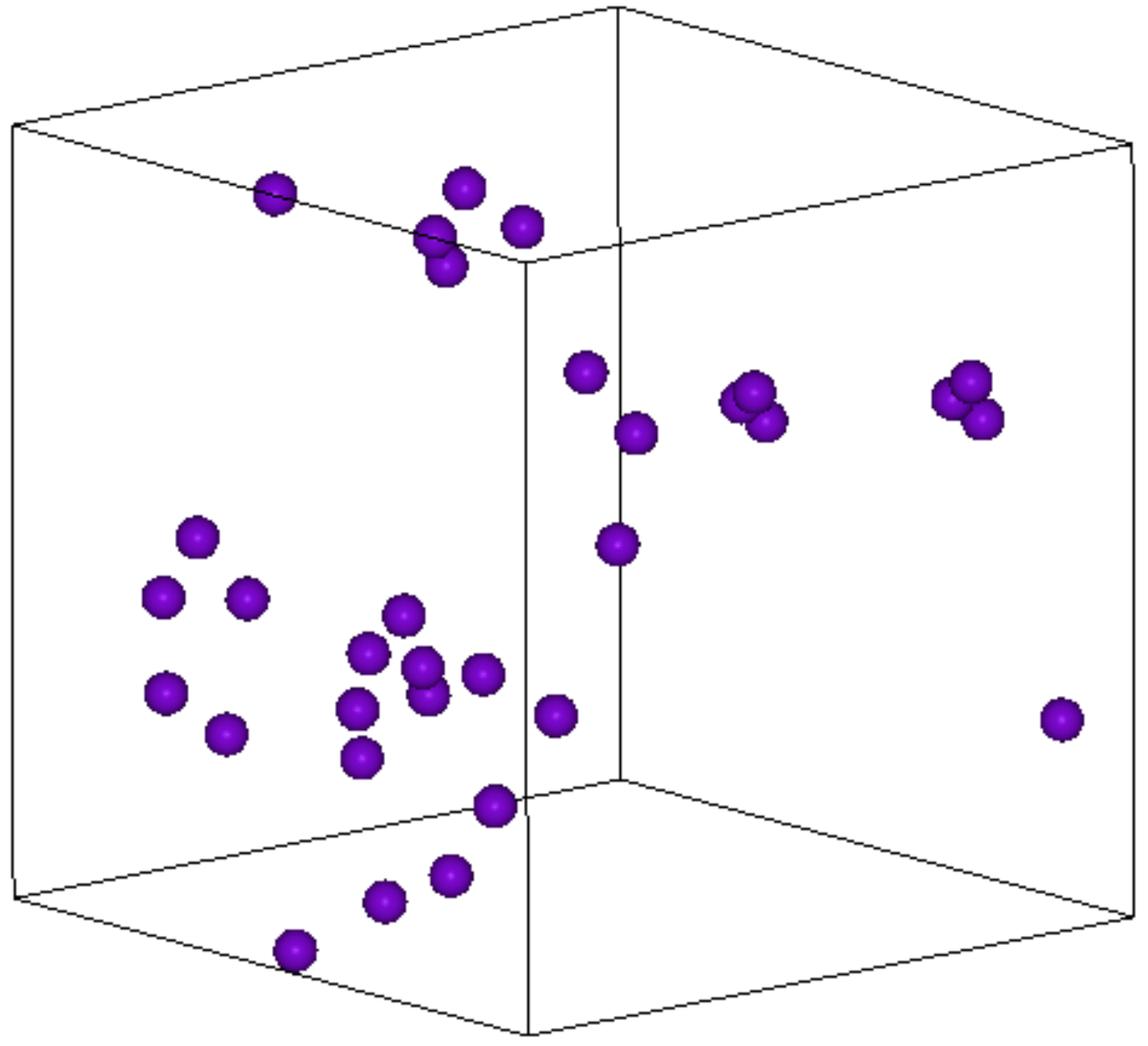}}
\subfigure[\label{fig1d} $C_1, C_0$]{\includegraphics[scale=0.25]{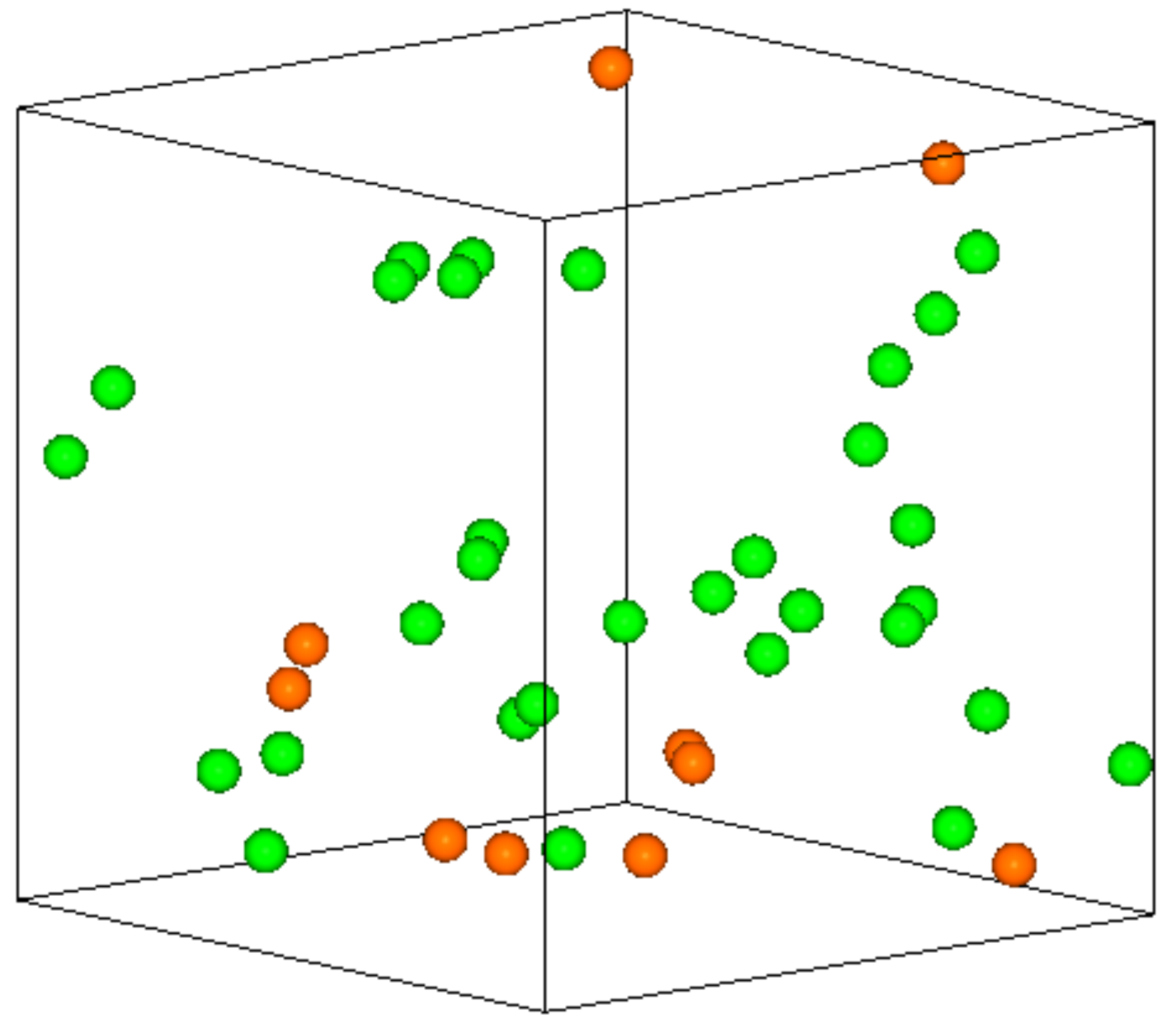}}
\caption{{\label{fig1}}{A particular disorder realization for $L=6$ and $x=1/32$. Colors at different sites in (a) and (b) represent different acceptance ratios of spin flips ($R_i$) using a conventional single spin-flip Metropolis algorithm. Sites with $(0 < R < 10^{-4})$ are denoted by violet, $(10^{-4}< R < 10^{-3})$ by blue, $(10^{-3}<R<10^{-2})$ by green, $(10^{-2}<R<0.1)$ by yellow, $(0.1 <R<0.25)$ by orange and $(0.25<R<1)$ by red. With the chosen cluster parameters ($a_s=1.3125$, $b_s=0.75$ and $C_L=N/5$), three cluster sets $C_0$, $C_1$ and $C_2$ are obtained for this disorder realization. (c) shows the member sites of the clusters that belong to the set $C_2$ (in violet). (d) shows the additional member sites of the clusters in $C_1$ that are already not part of $C_2$ (in green) and the additional member sites of the clusters in $C_0$ that are already not part of $C_2$, $C_1$(in orange). The figures were generated using the graphics software QMGA~\cite{software}.}
 }
\end{figure*}
%***************************************
%%%%Dynamic heterogeneity 
{\underline{\it Dynamic heterogeneity:}}
Monte-Carlo simulations with a single-spin flip Metropolis algorithm in 
combination with parallel tempering in temperature~\cite{HukushimaN96} 
is the method of choice to simulate 
Edwards-Anderson type models~\cite{EAnumerics1, EAnumerics2}. 
However, this local algorithm fails 
to equilibrate the Ising system considered in Eq.~\ref{eq:Ising} 
because of long autocorrelation times except for very small system 
sizes when $x \ll 1$. 
Apart from the computational effort scaling 
as $\mathcal{O}(x^2L^6)$ due to the long-ranged interactions, 
the other more serious bottleneck to equilibration is the 
presence of clusters of 
effectively frozen spins at low temperature under a single spin-flip 
dynamics. Their presence can be 
seen by monitoring the acceptance ratio, $R_i$, of the spin flips at 
each site $i$ in a particular disorder realization by 
performing a simulation using the Metropolis algorithm 
(without any parallel tempering in temperature). A disorder realization is 
produced by placing (ghost) spins on a fraction $x$ of sites that are randomly 
selected~\cite{comment1} from the $16L^3$ sites of the system of 
linear dimension $L$ (with $16$ sites in the conventional cubic unit cell of 
the pyrochlore lattice). Fig.~\ref{fig1a} and Fig.~\ref{fig1b} show 
the spatial distribution of $R_i$ for a particular disorder realization 
at $L=6$, $x=1/32$ at two different temperatures, $T=0.072 (\approx 2T_x)$ 
and $T=0.047 (\approx 1.3T_x)$, respectively. The data for $R_i$ has been 
obtained by averaging over $2000$ different runs where the different 
initial configurations at a temperature $T$  were equilibrated using 
our cluster algorithm, after which $10^6$ spin-flip attempts 
(per spin) were made using a Metropolis algorithm. 
A strong dynamic heterogeneity in the behavior of $R_i$  
is visible at both temperatures. The spins have a wide range of $R_i$ 
with several spins remaining practically frozen (violet and blue sites 
in Fig.~\ref{fig1a}, Fig.~\ref{fig1b}), others having an intermediate 
$R_i$ (green and yellow sites in Fig.~\ref{fig1a}, Fig.~\ref{fig1b}), 
and the rest having a high $R_i$ (orange and red sites in Fig.~\ref{fig1a}, 
Fig.~\ref{fig1b}). These effectively frozen spin clusters make the Metropolis 
algorithm highly inefficient for such dilute dipolar systems. Parallel 
tempering in temperature also fails to equilibrate such systems since the 
clustering effects persist even at temperatures like $T \approx 2T_x$ 
(Fig.~\ref{fig1a}) and above. Such clustering was also observed previously 
in a numerical study of the dynamics of uniaxial Ising spins~\cite{BiltmoH12} 
interacting via dipolar interactions in a dilute system.   

\begin{figure}[!]
\centering
\subfigure[{\label{fig2a}}]{\includegraphics[width=0.8\hsize]{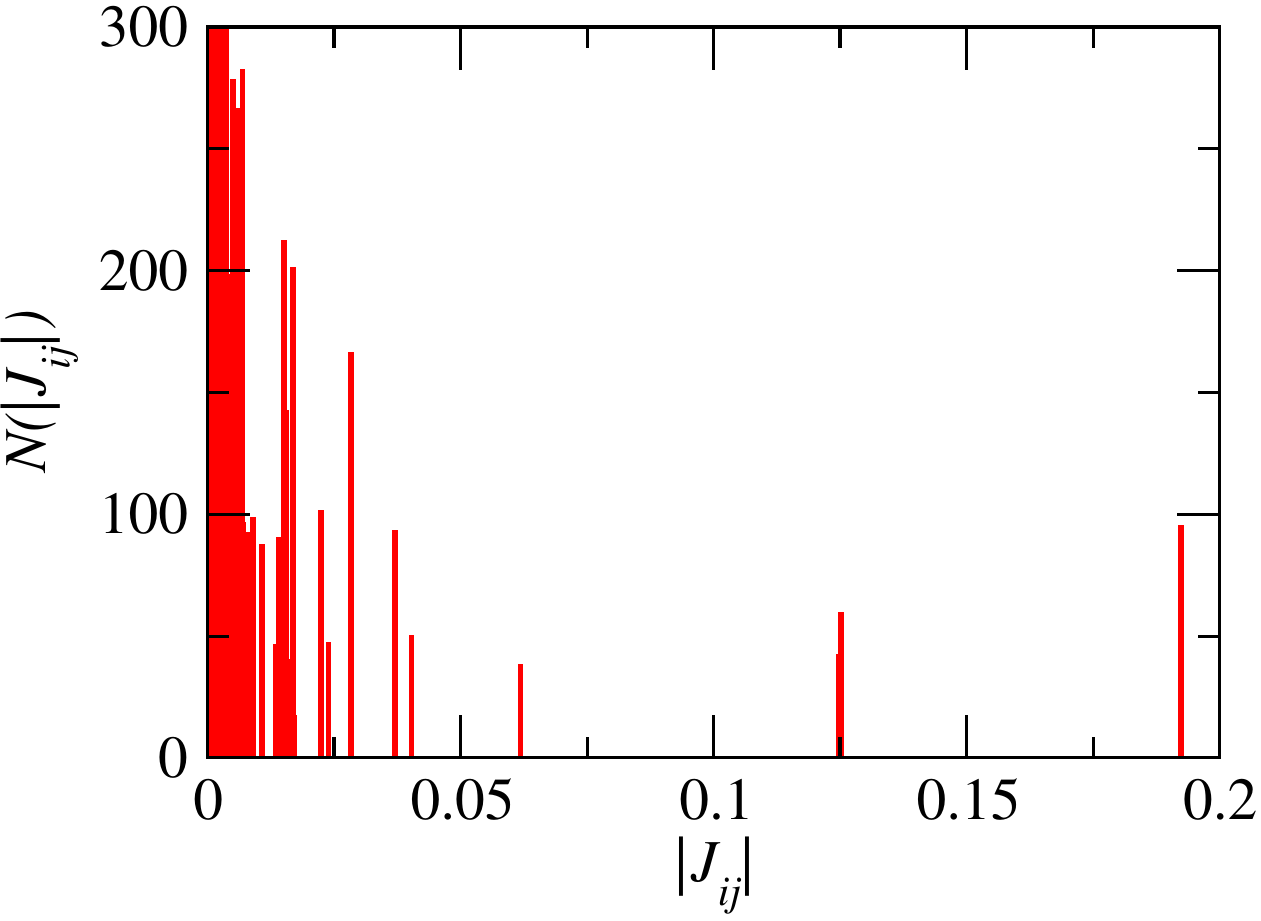}}\\
\subfigure[{\label{fig2b}}]{\includegraphics[width=0.7\hsize]{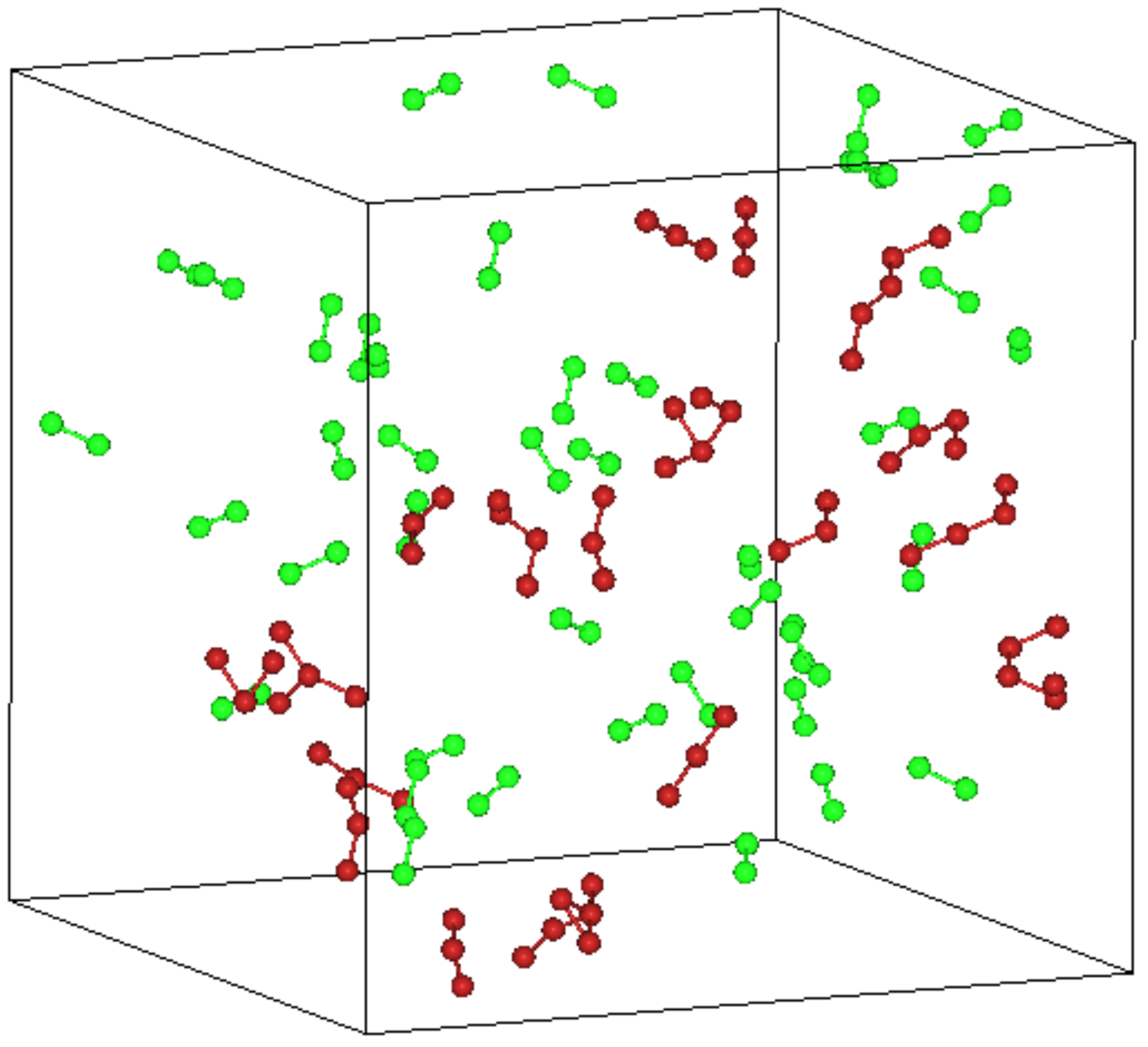}}%{Cn_snapshot.pdf}}
\caption{ {\label{fig2}}(a) Variation of the number of bonds, $N(|J_{ij}|)$, with $|J_{ij}|$ for a particular disorder realization at $L=10$ and $x=1/32$. While $J_{av} \sim 10^{-3}$ here, the maximum value of $|J_{ij}| \sim 0.2$. The y axis is truncated at $300$ for clarity. 
(b) The clusters in the set $C_n$ (constructed with the largest $J_T$) for a particular disorder realization at $L=10$ and $x=1/32$. The cluster construction parameters used were $a_s=1.3125$, $b_s=0.75$ and $C_L=N/5$. 
While the majority of the clusters in $C_n$ have size-$2$ (indicated in green), there are $n$-spin clusters with $n>2$ as well (indicated in red). Long bonds due to periodic boundary conditions have not been shown here.}
\end{figure}

%%%%.  Cluster algorithm 
{\underline{\it Cluster algorithm:}}
To ameliorate the slow equilibration due to these spin clusters, 
we present a modified version of a cluster algorithm 
used in Ref.~\onlinecite{AndresenKOS14}
(see also Refs.~\onlinecite{JanzenHE08, AndresenJK11}). 
The key idea is to incorporate correlated multi-spin flips to deal with the 
dynamically frozen spin clusters. The emergence of these clusters can 
be understood as follows: 
While the average distance between spins scale as $r_{av} \sim a x^{-1/3}$ and 
therefore the average of the magnitude of $|J_{ij}|\sim Dr_{av}^{-3} \sim Dx$ 
(the average value of $|J_{ij}|$ in a disorder realization is denoted by 
$J_{av}$ henceforth), which is also the reason behind the expectation 
that $T_x \propto x$ at small $x$, 
the minimum distance between the spins is fixed by the 
lattice constant $a$ and is independent of $x$. Therefore, in any given 
disorder realization at small $x$, there will be spin pairs $(i,j)$ such that 
$|J_{ij}|$ is much greater than $J_{av}$. Fig.~\ref{fig2a} shows the 
distribution of $|J_{ij}|$ in one disorder realization for $L=10$ at $x=1/32$. 
While $J_{av} \sim 10^{-3}$ in this case, there are several spin pairs 
for which $|J_{ij}| \gg J_{av}$ with the maximum 
value of $|J_{ij}| \sim 0.2$. These {\it tightly bound} 
spin pairs will be  effectively frozen
under a local single-spin flip Metropolis update at temperatures 
$T \sim \mathcal{O}(T_x)$ wherever $|J_{ij}| \gg T$. 
Furthermore, the wide distribution in the values of 
$|J_{ij}|$ at small $x$ (Fig.~\ref{fig2a}) due to the power-law nature of the 
interactions explains the wide spread in the values of $R_i$ as 
seen in Fig.~\ref{fig1a} and Fig.~\ref{fig1b} under a local single-spin 
flip dynamics. 

An additional correlated flip of these 
spin pairs (apart from the usual single-spin flips) satisfying 
detailed balance may seem to be the cure for this problem.   
However, there will also be frozen clusters in the system which are bigger 
than size-$2$ (Fig.~\ref{fig2b}) and cannot be handled 
by these pair flips alone. 
Consider any subset of these tightly bound spin pairs that 
form a {\it connected cluster} (Fig.~\ref{fig2b}) such that it is possible 
to get from every site in that cluster to every other site in it 
through these {\it strong} bonds, where a strong bond is set by the condition that $|J_{ij}| \gg T$, 
then all these spins in the cluster are 
mutually frozen as well with respect to single-spin flips at $T$. 
Ignoring the rest of the {\it weak} bonds in the 
system effectively breaks it into these connected clusters of spins. 
This suggests an immediate low-energy move where all the Ising spins 
$\{ S_i \}$ that belong to a cluster are flipped together to  $\{ -S_i \}$ 
irrespective of the values of these spins relative to each other. In 
Ref.~\onlinecite{AndresenKOS14}, the clusters were chosen to be {\it fully 
connected} such that all the $n(n-1)/2$ bonds between the spins of a 
$n$-spin cluster are strong bonds. However, consider a case where two (or more) 
fully connected clusters share one or more sites (e.g., two size-$2$ clusters 
formed by sites $(i,j)$ and $(j,k)$ share a common spin at site $j$ 
but with $|J_{ik}|$ small enough to be a weak bond). 
Then flipping all the spins of one such fully 
connected cluster would not necessarily be a low-energy move since it will 
only flip a subset of spins of the other one(s). 
To remedy this, one simply needs to flip all the member spins of 
these fully connected clusters that share the common site(s) simultaneously 
but this is the same as flipping a single connected cluster in our approach. 

%%%%%%%%%%%%%%%%%%%%%Fig. 2%%%%%%%%%%%%%%%%
%%%%%%%%%%%%%%%%%%%%%%%%%%%%%%%%%%%%%%%%%%%%%%
 
Our cluster construction procedure requires specifying three parameters 
$a_s$, $b_s$ and $C_L$. We then generate different sets of spin clusters 
for every disorder realization at the beginning of the simulation. 
We select all the bonds $(i,j)$ where $|J_{ij}| \geq$ \( a_s J_{av}\) 
and prepare a list $[(i,j)]$ of the bonds in which their strengths 
are arranged in ascending order of their 
magnitudes. We consider the smallest value of $|J_{ij}|$ from this list as 
$J_s$ and initially set $J_T=J_s$, where $J_T$ is a given target energy. 
We then search for all the bonds $(i,j)$ 
from the list $[(i,j)]$ 
such that \( |J_{ij}|\geq J_T\) and group them into connected 
clusters. The clusters contain only the site indices and a cluster 
move is the simultaneous flipping of all its Ising spins $S_i$ to $-S_i$. 
Therefore, the flipping of spins during the simulation does not change the 
definition of the clusters. The collection of clusters for a given $J_T$ 
forms a cluster set. If the size of the largest cluster (i.e., the number of 
its member spins) in the set 
exceeds a threshold size \(C_{L}\), we reject this cluster set and go to 
the next higher value in the list $[(i,j)]$ using: 
\(J_{T} \geq J_{T} + \Delta J \) where \(\Delta J = b_s J_s\) which we then 
use to generate a new set of clusters 
and again check the size of the largest cluster 
in it. In this way we sequentially build and reject the sets of 
clusters until we find a set in which the size of the largest cluster 
is \(\leqslant C_{L}\). We call this set ``\( C_{0}\)". 
After the formation of ``\( C_{0}\)", we form the next set ``\( C_{1}\)" 
by using only the bonds that have \( |J_{ij}|\geq J_T\) where 
\(J_{T} \geq J_{T} + \Delta J \) 
(and $J_T$ is again taken from the list $[(i,j)]$). 
We continue to generate more cluster sets upto the last set of 
clusters ``\( C_{n}\)" in this manner.

The parameter `\(a_s\)' controls the starting 
point of cluster constructions 
for the set $C_0$. The requirement is simply to start the construction 
such that the first `test' set may have at least one cluster of size 
\(> C_{L}\). We consider the parameter `\(b_s\)' to ensure that 
two successive sets are sufficiently different from each other. 
\(C_{L}\) can be chosen according to the size of the largest frozen cluster 
(with respect to single spin flips) which clearly increases as the 
temperature or $x$ is lowered. 
For most of the simulations, we consider the following parameter values : 
\(a_s = 1.3125,b_s=0.75\) and \(C_{L}=N/5\) where \(N=16xL^3\) 
is the total number of ghost spins in the respective configurations. 
The thermalization timescale of the algorithm depends on the parameters 
$a_s,b_s,C_L$ but we leave their systematic optimization to a future study 
(for some discussion, see Appendix~\ref{appendixA}).

Each cluster set contains clusters of different sizes. Each set has a majority 
of size-$2$ clusters. However, even the final set \( C_{n}\) may consist 
of multiple clusters of size \(> 2\), especially at large $L$ 
(See Fig.~\ref{fig2b}). 
A small cluster in set \( C_{l}\) may well be a part of a larger cluster 
present in another set \( C_{m}\) where \(m < l\). 
A particular cluster may be a member of 
multiple sets.  
Going from a cluster set $C_m$ to a set $C_l$ where $m>l$ entails the 
following: 
(a) formation of new clusters not present in $C_m$, (b) growth of 
clusters contained in the set $C_m$, and (c) clusters in $C_m$ merging 
to form bigger clusters in $C_l$.

The multiple cluster sets $C_0$, $\cdots$, $C_n$, 
each with a different $J_T$, are 
constructed since the interaction $J_{ij}$ has a power law nature and 
thus each disorder realization has a hierarchy of energy scales 
(Fig.~\ref{fig2a}).  
The clusters in the set $C_n$ (that has the highest $J_T$) mimic the 
dynamically frozen spin clusters that are formed at 
higher temperatures (Fig.~\ref{fig1c})
whereas the clusters in $C_0$ mimic frozen spin clusters at 
lower temperatures (Fig.~\ref{fig1d}). 
The member spins of the clusters in set $C_n$ 
are typically composed of spins that have 
the lowest $R_i$ under a local single-spin flip 
Metropolis algorithm (see Fig.~\ref{fig1c}). New members of $C_{n-1}$ etc 
(which do not already belong to the previous sets $C_n$ etc) 
typically have progressively higher 
values of $R_i$ (but still much lower than the spins with the highest 
values of $R_i$ in the system) as can be seen from Fig.~\ref{fig1d}. 
Our cluster algorithm thus correctly identifies the majority of the 
frozen spins present in the system at small $x$ as well as the 
heterogeneity in their dynamic behavior (Fig.~\ref{fig1}) 
by associating them to different sets.

We need to include conventional single-spin flip moves in our cluster 
algorithm as well not only to keep the Monte Carlo dynamics ergodic 
(since there are spins which are not part of any cluster) 
but also for breaking the size-$2$ clusters which is only possible 
via single spin flips. Similarly, the clusters in a 
set $C_m$ are instrumental in breaking the bigger clusters 
in a set $C_l$ where $m>l$. 
During our simulation, at each step, we apply either 
a single spin flip move or a cluster flip move. The probability that a 
single spin flip is attempted is taken as $85\%$ and that a cluster flip move 
is attempted is then taken as $15\%$ in most of the simulations. 
In previous works~\cite{AndresenKOS14,JanzenHE08, AndresenJK11}, 
a cluster was randomly (uniformly) selected from all possible sets and then a 
cluster flip was attempted.  
In our approach, each cluster set is assigned a 
probability of being chosen during 
the cluster flip move which is taken to be non-uniform, with the highest  
(lowest) weight given to clusters in $C_n$ ($C_0$).
This way we ensure that the more strongly coupled spins are attempted 
to be flipped more often. 
 
%The clusters in set $C_n$ represent the strongest bottlenecks to 
%equilibration (given that the clusters in $C_n$ have the highest fraction of 
%the maximally frozen spins (Fig.~\ref{fig1})), followed by clusters in $C_{n-1}$ and so on. 
%The relative importance of the different cluster sets in achieving 
%equilibration is discussed further in the Appendix~\ref{appendixA}.

Specifically, we select the set \( C_{n}\) with probability 
\(P_{n}=1/2\), the probability that we select `k-th' set \( C_{k}\) is 
$P_{k}=(1-\sum\limits_{i=k+1}^{n}P_{i})/2$. The probability that we select 
the set \( C_{0}\) is then \(P_{0}=\)\((1-\sum\limits_{i=1}^{n} P_{i})\). 
Once a particular cluster set is chosen, then all the clusters of that set 
have equal probability to be chosen for the actual cluster flip attempt.
The relative importance of the different cluster sets in achieving 
equilibration is discussed further in the Appendix~\ref{appendixB}.

Both the single-spin flips and cluster flips are accepted with the Metropolis 
probability min$\left[1, \exp(-\Delta E/T )\right]$, where \(\Delta E\) 
is the energy difference between the new configuration and the old 
configuration, to preserve detailed balance. One Monte-Carlo step (MCS)
consists of \(N\) spin/cluster flip attempts in total. 
We further use parallel tempering in temperature~\cite{HukushimaN96} 
in combination with the cluster algorithm to accelerate equilibration. 
In detail, we simulate \(2N_{T}\) replicas at $N_T$ different temperatures 
in parallel (thus, two independent replicas at each temperature so that we 
can calculate overlap observables defined in Eq.~\ref{SGop}), 
with the consecutive temperatures scaled by a factor \(c\) 
such that \(T_{n} = (1+c)^{n} T_{0}\), where 
\(n = 0,1, 2,\cdots, N_{T}-1\)~\cite{Gingras}. 
The parameters \(T_{0}, c \mbox{ and } N_{T}\) are adjusted so that 
the acceptance ratio for parallel tempering swaps between neighbouring 
temperatures is \(> 50\%\). For the exchange process, the replica pairs 
\((T_{m}, T_{m+1})\) are divided into two subgroups, i.e., odd-\(m\) and 
even-\(m\) groups. The exchange trial is performed for one of these 
subgroups after every 20 MCS. This sequence
which we denote as a Monte Carlo Sweep (MC Sweep) 
is repeated several times during the course of 
the simulation. A large number of independent disorder realizations 
(denoted by $N_{sample}$, which is $1500$ or more) 
are taken to perform the disorder averaging. 
%%%%%%%%%%%%%%%%%%%%%%%%%%%%%%%%%%%%%%%%%%%%%

{\underline{\it Observables:}}
Let us now describe the observables. The spin glass order parameter, 
$q_{EA}^{\alpha \beta}(\mathbf{k})$, at wavevector $\mathbf{k}$ is defined as
%%%%%%%%%%%%%%%%%%%%%%%%%%%%%%%
\begin{eqnarray} 
q_{EA}^{\alpha \beta}(\mathbf{k}) = \frac{1}{N}\sum_i \mu_i^{\alpha (1)} \mu_i^{\beta (2) }\exp(i\mathbf{k} \cdot \mathbf{r}_i) 
\label{SGop}
\end{eqnarray}
%%%%%%%%%%%%%%%%%%%%%%%%%%%%%%%
where $\alpha,\beta=x,y,z$ are the spin components (where the ghost spins 
point along the local easy axes) and $(1)$ and $(2)$ denote two identical 
disorder realizations of the system with the same set of interactions. 
From this, we calculate the spin glass susceptibility, 
$\chi_{SG}(\mathbf{k})$, defined as 
%%%%%%%%%%%%%%%%%%%%%%%%%%%%%%%
\begin{eqnarray}
\chi_{SG}(\mathbf{k}) = N \sum_{\alpha,\beta}[\langle|q_{EA}^{\alpha \beta}(\mathbf{k})|^2 \rangle] 
\label{SGchi}
\end{eqnarray} 
%%%%%%%%%%%%%%%%%%%%%%%%%%%%%%%
where $\langle \cdots \rangle$ and $[\cdots]$ denote thermal and disorder 
averages, respectively. In particular, $\chi_{SG} \equiv 
\chi_{SG}(\mathbf{k}=\mathbf{0})$ is an indicator for the spin glass 
transition since above (below) the transition, $\chi_{SG}$ is 
finite (diverges) as $L \rightarrow \infty$. Furthermore, a spin glass 
correlation length $\xi$ can also be defined by using the following relation:
%%%%%%%%%%%%%%%%%%%%%%%%%%%%%%%
\begin{equation}
 \xi=\frac{1}{2\sin \left(\frac{|\mathbf{k}_{\rm min}|}{2} \right)} \left(\frac{\chi(\mathbf{k}=\mathbf{0})}{\chi(\mathbf{k}_{\rm min})}-1 \right)^{\frac{1}{2}}
\label{SGxi}
\end{equation}
%%%%%%%%%%%%%%%%%%%%%%%%%%%%%%%
where $\mathbf{k}_{\rm min} = \frac{2\pi}{L}(1,0,0)$. The ratio $\xi/L$ 
approaches a universal value characteristic of the critical point 
as $L \rightarrow \infty$ in case the spin glass transition is 
continuous in nature.  
%%%%%%%%%%%%%%%%%%%%%%%%%%%%

\begin{figure}[!]
\centering
\subfigure[{\label{fig3a}}]{\includegraphics[scale=0.3]{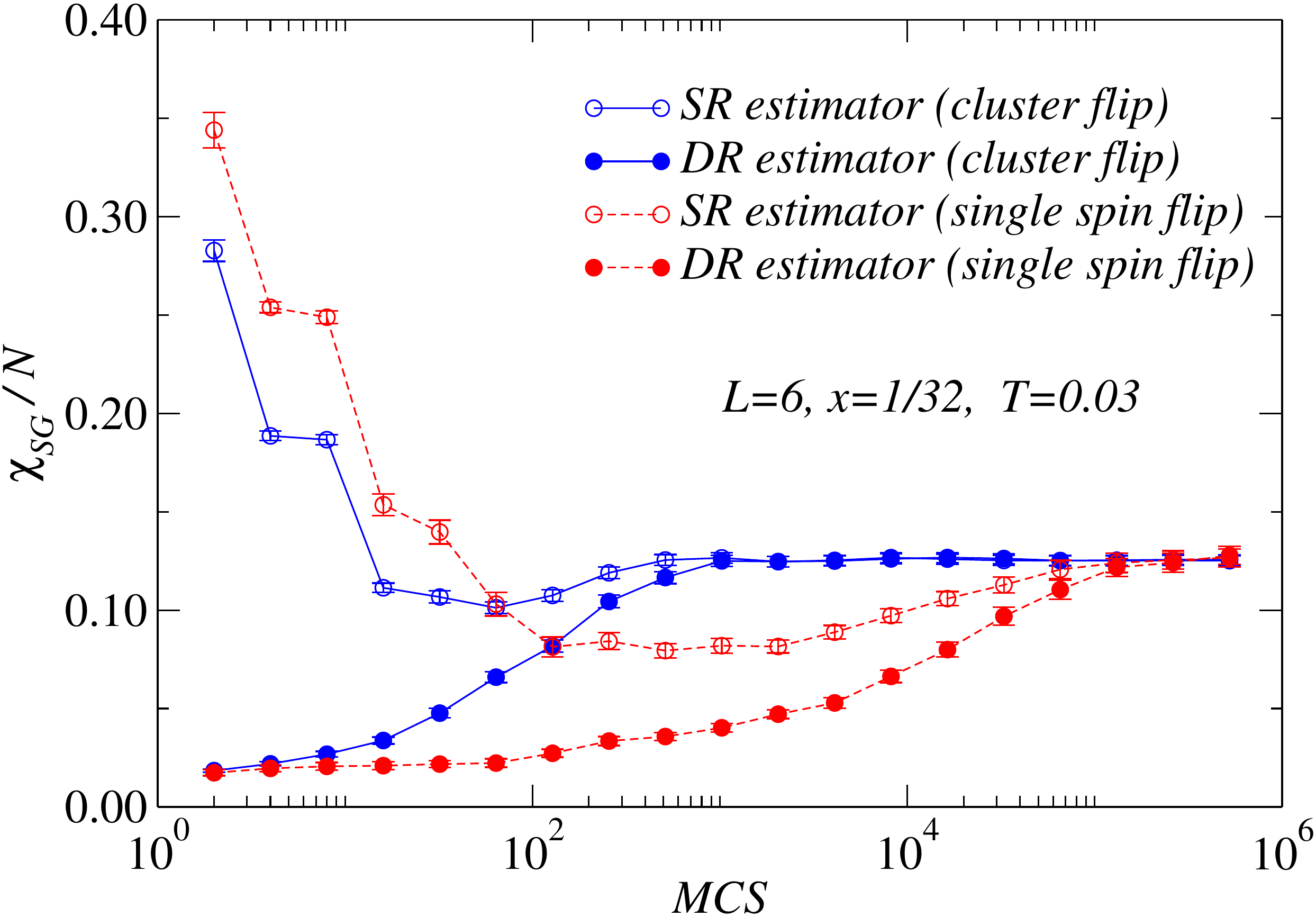}}\\
\subfigure[{\label{fig3b}}]{\includegraphics[scale=0.3]{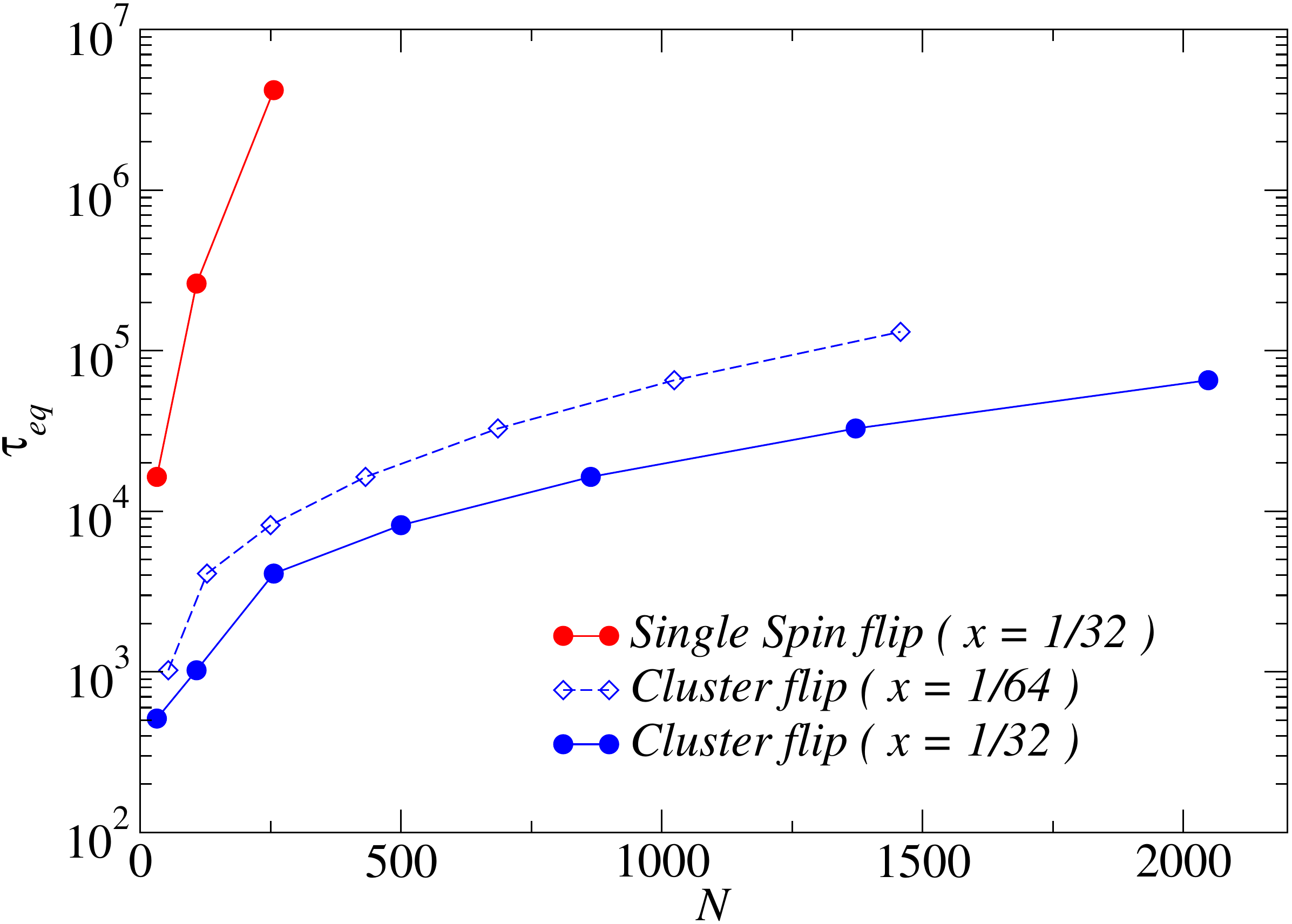}}
\caption{ {\label{fig3}}(a) Time variation of $\chi_{SG}/N$ calculated from both the double replica (DR) and the single replica (SR) estimators of the spin overlap function as defined in Eq.~\ref{equiltest} using both the cluster algorithm (denoted by cluster flip) and the single spin flip algorithm (denoted by single spin flip) in combination with parallel tempering in temperature. The autocorrelation time $\tau_{eq}$ estimated from such an analysis is shown in (b) for various $N$ at $x=1/32$ and $x=1/64$.}
\end{figure}

{\underline{\it Equilibration test and autocorrelation time analysis:}}
To test the equilibration of the algorithm, we measure 
$q_{EA}^{\alpha \beta}(\mathbf{k}=\mathbf{0})$ using a double replica 
(DR) (Eq.~\ref{SGop}) and a single replica (SR) estimator and 
calculate $\chi_{SG}$ using Eq.~\ref{SGchi}. The estimators are as follows:
%%%%%%%%%%%%%%%%%%%%%%%%%%%%%%%%%%%%%%
\begin{subequations}
%\begin{equation}
\begin{alignat}{2}
q_{DR}^{\alpha \beta}(t_0) &= \frac{1}{Nt_0}\sum_{t=1}^{t_0}\sum_i \mu_i^{\alpha (1)} (t_0+t)\mu_i^{\beta (2) } (t_0+t) \\
%\end{equation}
%\begin{equation}
q_{SR}^{\alpha \beta}(t_0) &= \frac{1}{Nt_0}\sum_{t=1}^{t_0}\sum_i \mu_i^{\alpha (1)} (t_0+t)\mu_i^{\beta (1) } (2t_0+t) 
%\end{equation}
\end{alignat}
\label{equiltest}
\end{subequations}
%%%%%%%%%%%%%%%%%%%%%%%%%%%%%%%%%%%%%
where each time step denotes a MCS and $t_0=2^n$ where $n=1,2,3, \cdots$. 
The DR (SR) estimator for the spin glass susceptibility at 
$\mathbf{k}=\mathbf{0}$ is then calculated using 
$q_{DR(SR)}^{\alpha \beta}(t_0)$ in Eq.~\ref{SGchi} 
and averaging over $300$ disorder realizations. 
For the initial condition, the two replicas for each disorder realization 
are taken to be in uncorrelated random spin configurations due to 
which $\chi_{DR}(t_0) \sim \mathcal{O}(1)$ while $\chi_{SR}(t_0) \sim 
\mathcal{O}(N)$ at small $t_0$. At sufficiently large $t_0$, determined by the 
autocorrelation time of the algorithm $\tau_{eq}$, both the estimators should 
converge to the correct equilibrium value after which it becomes 
independent of $t_0$ (within statistical errors)~\cite{BhattY88}. 
We show the results obtained as a function of 
$t_0 (=2^n)$ using both the single-spin flip algorithm and the 
cluster algorithm in combination with parallel tempering in Fig.~\ref{fig3a} 
for $L=6$, $x=1/32$ at a low temperature of $T=0.03$. 
From the data, it is clear that even for such a small system size, 
the cluster algorithm provides a reduction of the autocorrelation time 
(in units of MCS) 
by a factor of around $256$ as compared to the single-spin flip algorithm. 
We plot the autocorrelation times $\tau_{eq}$ estimated using both the 
algorithms as a function of $N$ at two different $x=1/32, 1/64$ in 
Fig.~\ref{fig3b}. For $x=1/32(1/64)$, we take $T_0=0.030(0.015)$ 
(since $J_{av}\sim Dx$), $c=0.065$ and $N_T=15$ for parallel tempering 
and show the results at the lowest temperature $T_0$ for both $x$.

Firstly, we notice the rapid growth of equilibration time by nearly a 
factor of $4000$ when $N$ increases from $32$ ($L=4$) 
to $256$ ($L=8$) at $x=1/32$ using the 
single-spin flip algorithm. For a smaller $x=1/64$, 
the equilibration time is $>10^6$ MCS even for a small size of $L=6$ ($N=54$). 
On the other hand, the equilibration time increases much more slowly with 
increasing $N$ and decreasing $x$ for the cluster algorithm. 
Note that since we do not change the parallel tempering parameters 
with system size for obtaining the results in Fig.~\ref{fig3b}, 
the acceptance ratio of the parallel tempering swaps is only around 
$12\%$ for $N=2048$ ($L=16$ at $x=1/32$), in spite of which the cluster 
algorithm manages to equilibrate the system.   
%%%%%%%%%%%%%%%%%%%%%Fig. 2%%%%%%%%%%%%%%%%
%%%%%%%%%%%%%%%%%%%%%%%%%%%%%%%%%%%%%%%%%%%%%%

\begin{figure}[!]
\centering
\subfigure[{\label{fig4a}} ]{\includegraphics[scale=0.3]{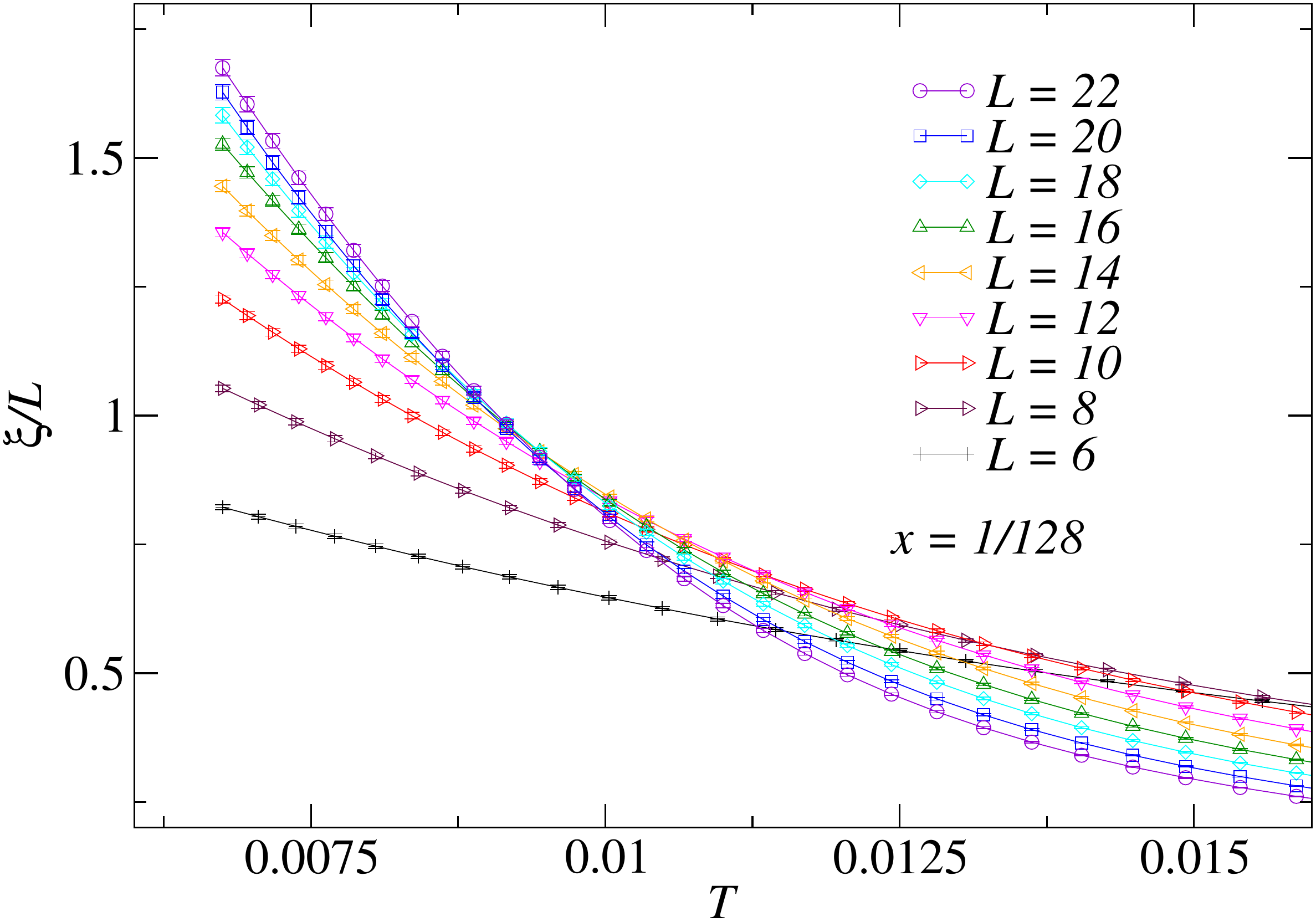}}\\
\subfigure[{\label{fig4b}} ]{\includegraphics[scale=0.3]{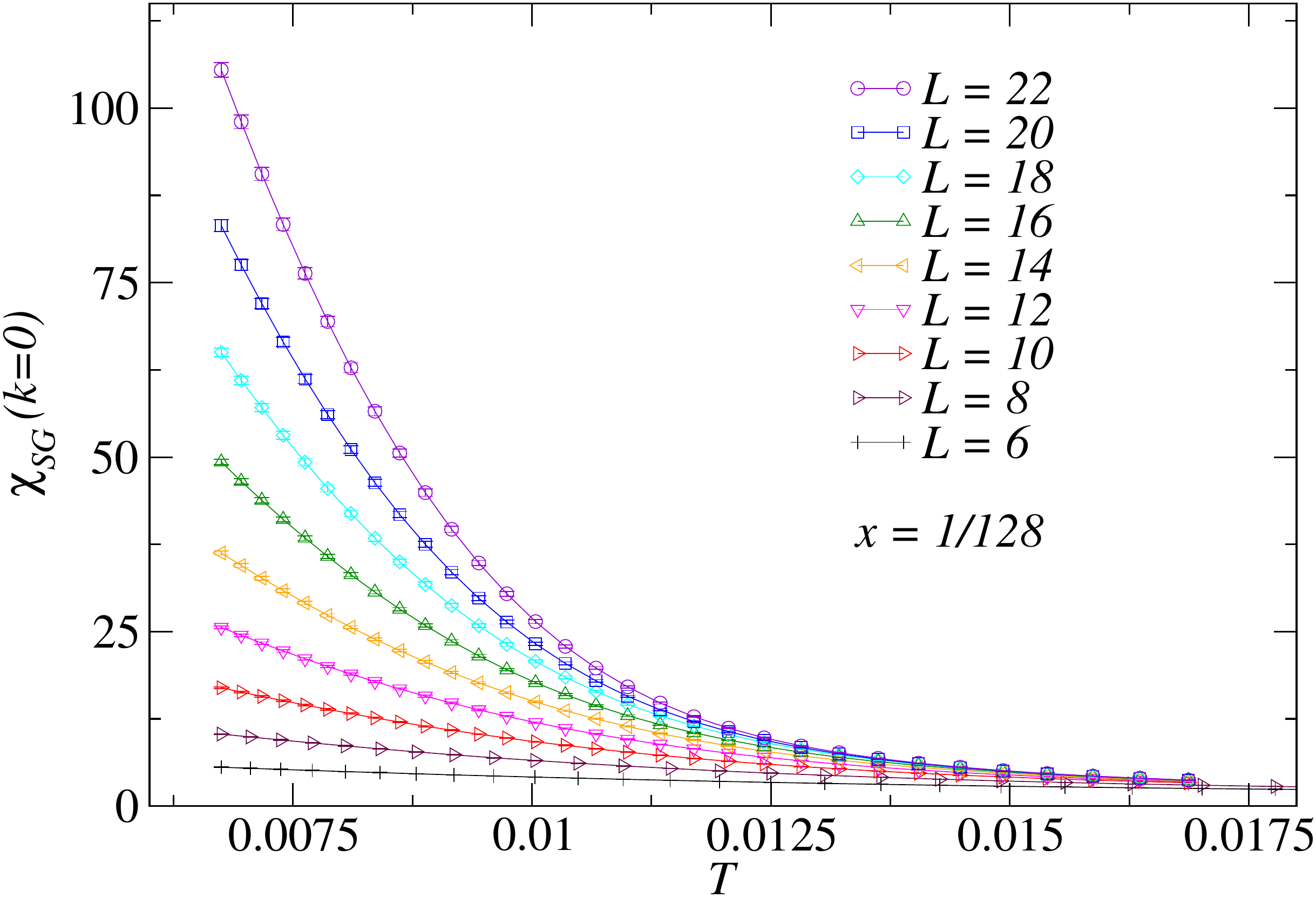}}
\caption{\label{fig4}{The behavior of (a) $\xi/L$ and (b) $\chi_{SG}$ as a function of $T$ shown for various $L$ at $x=1/128$. }}
\end{figure}

{\underline{\it Results:}}
Using our improved cluster algorithm, we study the behavior of 
$\chi_{SG}$ (Eq.~\ref{SGchi}) and $\xi/L$ (Eq.~\ref{SGxi}) 
in the following ranges: $4 \leq L \leq 16$ for $x=1/32$, 
$4 \leq L \leq 20$ for $x=1/64$, and $6 \leq L \leq 22$ for $x=1/128$ 
to understand the spin-glass transition at small $x$.  
The details regarding the simulation parameters are given in 
Appendix~\ref{appendixC}. 
We check for the proper thermalization of these quantities by using a 
standard logarithmic binning analysis, where the different observables 
are  calculated  by  using  data  only  from the second half of the 
measurements, the second quarter of them, the second eighth of them and 
so on. Equilibration is reliably achieved when at least the last 
three bins agree within error bars. The behavior of $\chi_{SG}$ and $\xi/L$ 
as a function of $T$ for different linear dimensions $L$ is shown in 
Fig.~\ref{fig4a} and Fig.~\ref{fig4b} for $x=1/128$ which strongly 
suggests a transition to a spin-glass phase as the temperature is lowered. 

%***********************************
%%%%%%%%%%%%%%%%%%%%%%%%%%%%%%%%%%%%%%%%%%

%%%%%%%%%%%%%%%%%%%%%%%%Fig. 3%%%%%%%%%%%%%
\begin{figure}[!]
\centering
\subfigure[{\label{fig5a}} ]{\includegraphics[scale=0.3]{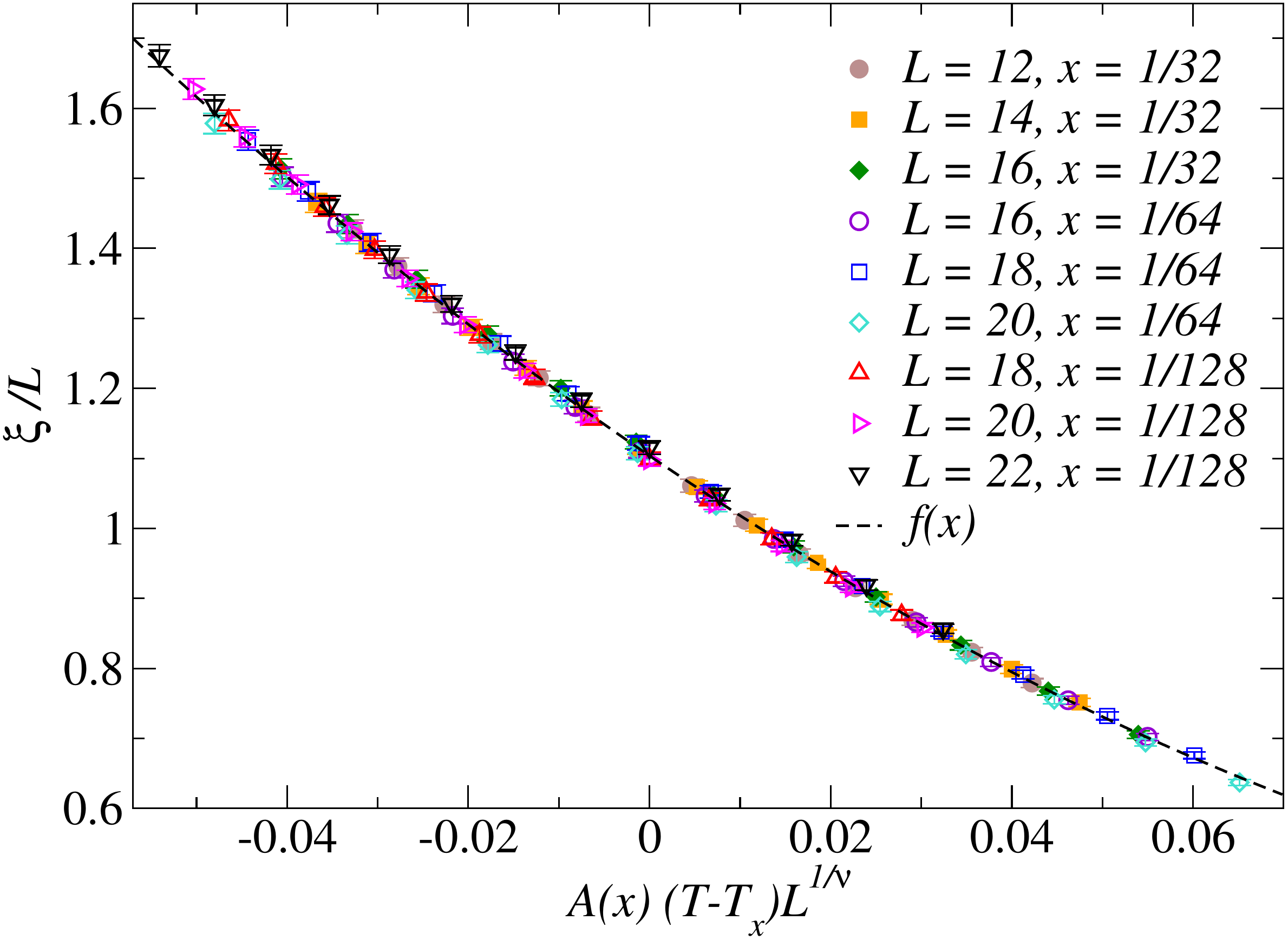}}\\
\subfigure[{\label{fig5b}} ]{\includegraphics[scale=0.3]{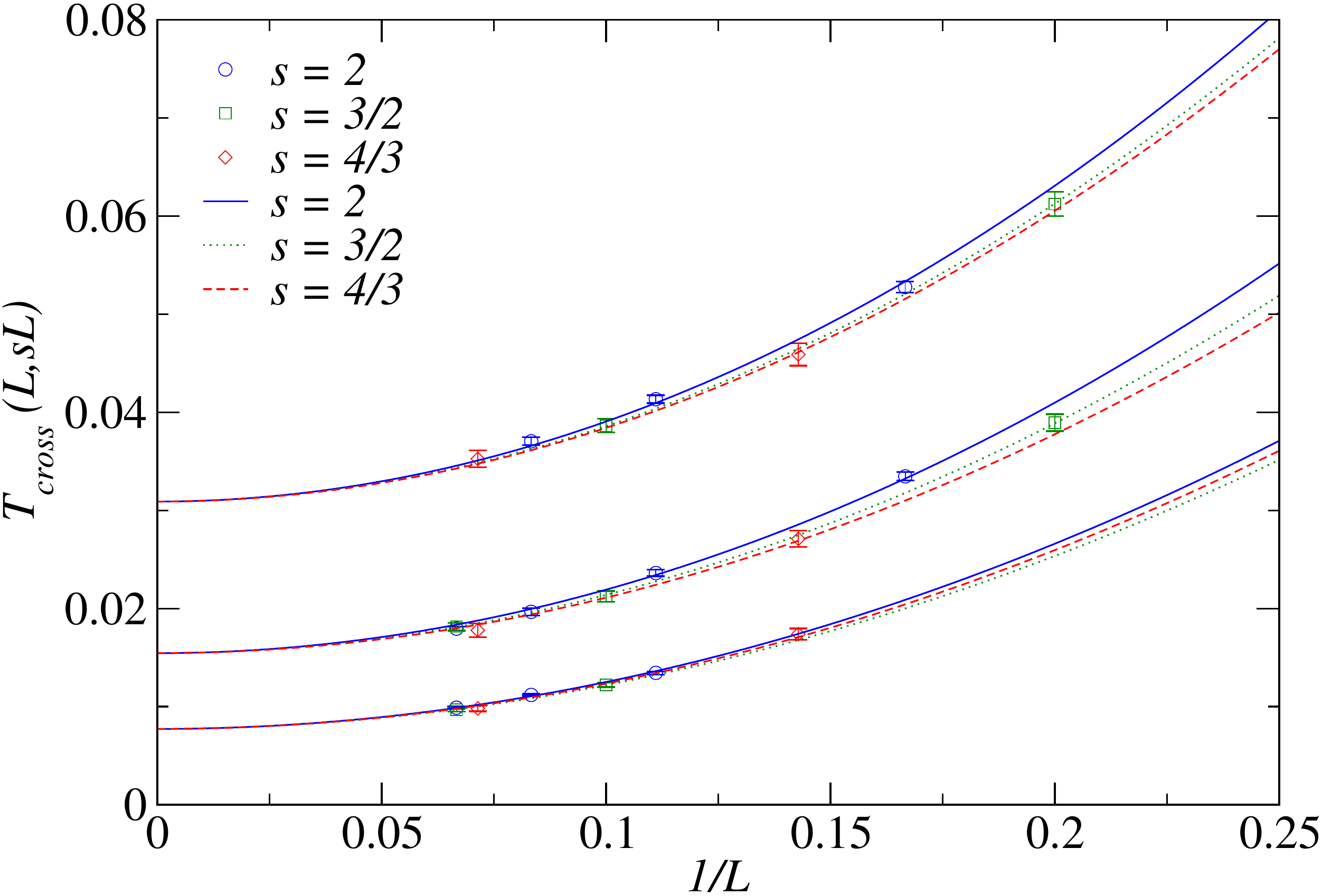}}
%\subfigure[{\label{fig5c}} ]{\includegraphics[scale=0.3]{Fixed_Tc_and_Fixed_nu.pdf}}
\caption{{\label{fig5}}(a) Scaling collapse of $\xi/L$ assuming the form in Eq.~\ref{FSS1}(a). (b) Behaviour of the crossing point $T_{cross}(L,sL)$ as a function of $1/L$ and its fit to the form shown in Eq.~\ref{Tcross}.}
\end{figure}
%%%%%%%%%%%%%%%%%%%%%%%%%%%%%%%%%%%%%%%%%%%%%%

%%%%%%%%%%%%%%%%%%%%%%%%%%%%%%%%%%%%%%%%%%%%%%%%%%%
%***************************************
To extract the transition temperature $T_x$ and establish the 
universality class of the transition, we now discuss the finite-size 
scaling behavior of these two quantities. Our results give strong 
evidence that the critical points at small $x$ are identical upto a 
simple $x$-dependent global rescaling and thus have the same universal 
physics. Assuming this scenario, to leading order in finite size scaling, 
$\xi/L$ and $\chi_{SG}$ behave as follows~\cite{KatzgraberKY06}:
\begin{subequations}
%\begin{equation}
\begin{alignat}{2}
\xi/L &= \mathcal{F}_1 (A(x)(T-T_x)L^{1/\nu}) \\
%\end{equation}
%\begin{equation}
B(x)\chi_{SG} &= L^{2-\eta}\mathcal{F}_2 (A(x)(T-T_x)L^{1/\nu})
%\end{equation}
\end{alignat}
\label{FSS1}
\end{subequations}
where $\mathcal{F}_{1,2}$ are universal functions, $\nu$ and $\eta$ are 
exponents characterizing the continuous transition, $T_x$ is the critical 
temperature at $x$ and $A(x),B(x)$ are ``metric factors'' that depend only on 
$x$.

We first perform the scaling collapse for $\xi/L$ since it has a smaller number of fitting parameters 
(Eq.~\ref{FSS1}(a)). We assume $T_x =ax(1+bx)$ to see the importance of the 
nonlinear terms at small $x$. The data collapse of 
$\xi/L$ (Fig.~\ref{fig5a}) 
gives $a=1.10(2)$ and $b=0.62(15)$ which determines $T_x$ and the 
critical exponent $\nu=1.27(8)$ with a reduced chi square per degree of 
freedom $\bar{\chi}^2=1.14$ (see Eq.~\ref{chisq} for definition of 
$\bar{\chi}^2$). The metric factors are determined to 
be $A(x=1/64)=1.65(3)$ and $A(x=1/128)=2.77(6)$ keeping $A(x=1/32)=1$.
Our estimate of $\nu$ agrees with that of Ref.~\onlinecite{AndresenKOS14} 
for uniaxial dipolar Ising spins in the dilute limit.  
For completeness, we show the data collapse of $\xi/L$ at each individual 
$x$ in Fig.~\ref{fig7d}, Fig.~\ref{fig7e}, Fig.~\ref{fig7f} 
and give the extracted $T_x,\nu$ and $\bar{\chi}^2$ in Table.~\ref{tabIII}.

We also perform a crossing point analysis of $\xi/L$ between systems of 
linear dimension $L$ and $sL$ (with a fixed $s$) 
at each $x$ to extract $T_x$. The crossing temperature, 
$T_{\mathrm cross}(L,sL)$, should converge to $T_x$ as $L \rightarrow \infty$ 
in the following manner~\cite{FernandezMGTY09}: 
\begin{eqnarray}
T_{\mathrm cross}(L,sL) = T_x + A_{SG}(x,s)L^{-(1/\nu +\omega)}
\label{Tcross}
\end{eqnarray}
where $A_{SG}(x,s)$ is a (non-universal) constant that depends on both 
$x$ and $s$ and $\omega$ is the exponent for the leading correction to 
scaling. Since Fig.~\ref{fig5a} already shows strong evidence 
that the universality 
does not depend on $x$ (for small $x$), we therefore assume the 
combination $(1/\nu +\omega)$ to be independent of $x$ and use 
$s=2,3/2,4/3$ to obtain the crossing of $\xi/L$ curves for $(L,sL)$ for 
various $L$ at $x=1/32, 1/64,1/128$. We then fit all the crossing point 
data for the different $x$ simultaneously to Eq.~\ref{Tcross} by assuming 
$T_x \propto x$ and $A_{SG}(x,s)$ to be different constants depending on the 
values of $x$ and $s$ respectively. The result is shown in Fig.~\ref{fig5b} and 
yields $T_x =1.00(3)x$ which is in good agreement with the previously 
obtained value of $T_x$ (Fig.~\ref{fig5a}) from the scaling 
collapse of $\xi/L$. We also obtain $1/\nu +\omega = 1.98(16)$ from the fit.

%%%%%%%%%%%%%%%%%%%%%%%%Fig. 3%%%%%%%%%%%%%
\begin{figure}[!]
\centering
\subfigure[{\label{fig6a}} ]{\includegraphics[scale=0.3]{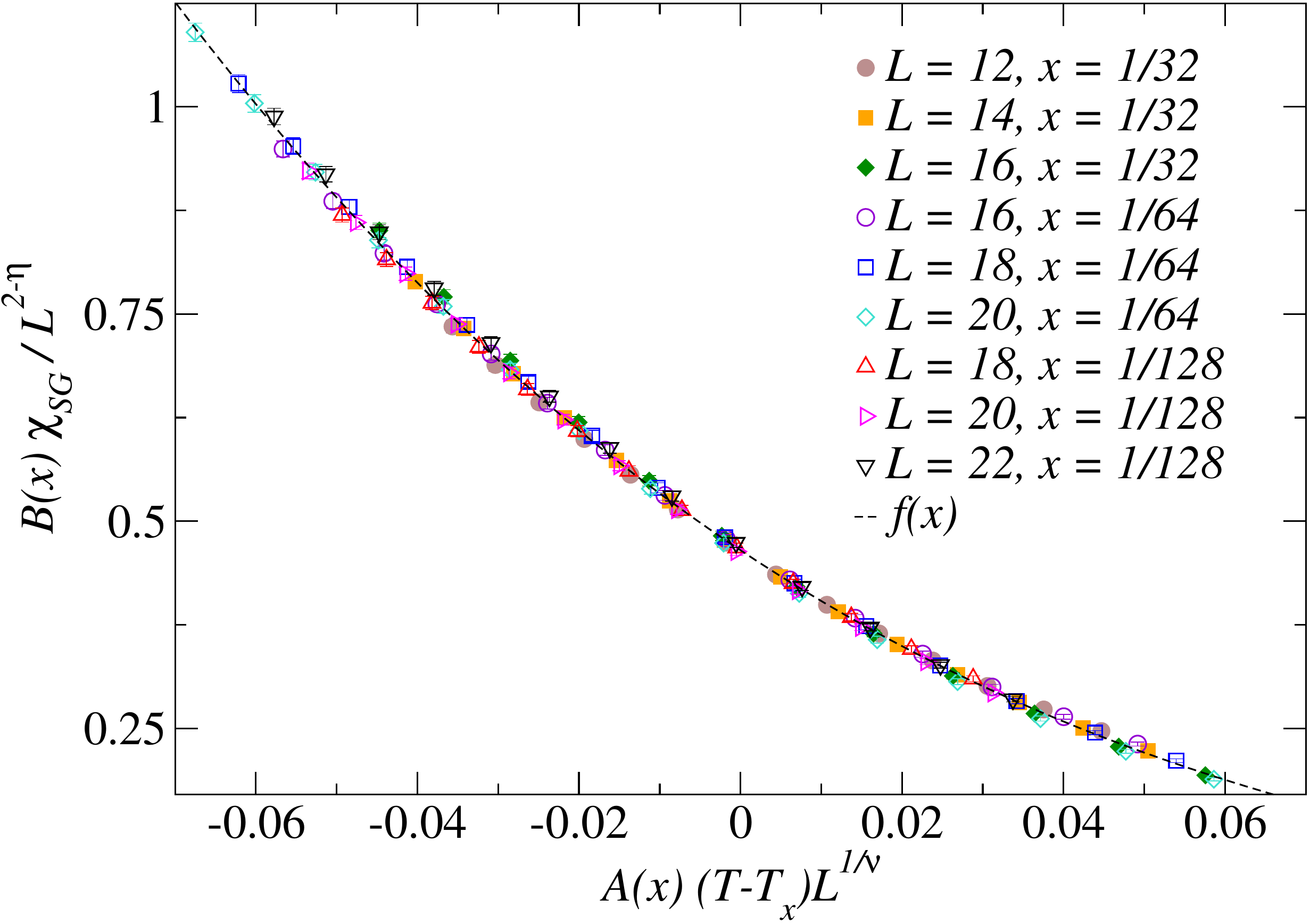}}\\
%\subfigure[{\label{fig6b}}]{\raisebox{.3cm}{\includegraphics[scale=0.55]{eta22.pdf}}}
\subfigure[{\label{fig6b}}]{\includegraphics[scale=0.3]{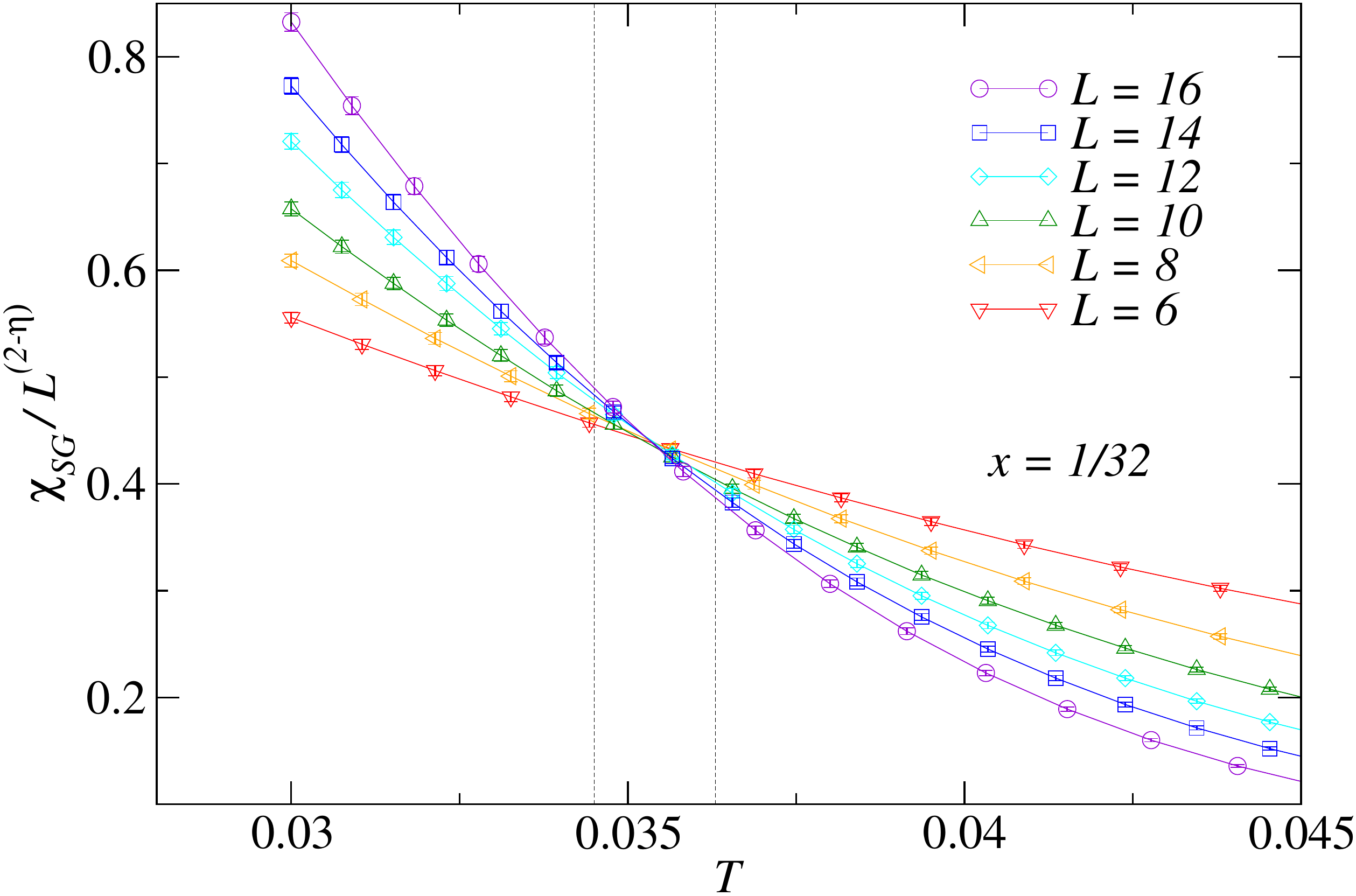}}
\caption{{\label{fig6}} (a) Scaling collapse of the spin glass susceptibility $\chi_{SG}$ assuming the form in Eq.~\ref{FSS1}(b) after fixing $T_x$ and $\nu$ from the collapse of $\xi/L$. (b) The behavior of $\chi_{SG}/L^{2-\eta}$ versus $T$ shown for $\eta=0.22$ at $x=1/32$. The dotted vertical lines indicate $T_x \pm \Delta T_x$ where $T_x$ is extracted from the data collapse of $\xi/L$ and $\Delta T_x$ is the statistical error bar.}
\end{figure}
%%%%%%%%%%%%%%%%%%%%%%%%%%%%%%%%%%%%%%%%%%%%%%

We now estimate the anomalous exponent $\eta$ from the behavior of 
$\chi_{SG}$. The exponent $\eta$ could not be reliably estimated for a 
dilute system of uniaxial dipolar Ising spins due to large finite-size 
corrections to scaling~\cite{Gingras, AndresenKOS14}. 
However, in the microscopic model adopted in this work 
(which provides a different lattice 
relization to the same universal physics), we can reliably extract $\eta$. 
The scaling collapse of $\chi_{SG}$ using Eq.~\ref{FSS1}(b)  
gives a large statistical error on the determination of $\eta=-0.35(69)$ 
when we keep $T_x(=ax(1+bx)),\nu,\eta,A(x),B(x)$ as free parameters for the 
fit. We, therefore, reduce the number of free parameters 
in the fit by fixing $T_x$ and $\nu$ 
from the previous data collapse of $\xi/L$. 
This gives us a much 
better estimate of $\eta=0.228(35)$ alongwith $A(x=1/64)=1.66(1), A(x=1/128)=
2.69(3), B(x=1/64)=1.52(2), B(x=1/128)=2.24(4)$ 
(with $A(x=1/32)=B(x=1/32)=1$) with $\bar{\chi}^2=1.78$ (see 
data collapse of $\chi_{SG}$ in Fig.~\ref{fig6a}). 
We note that the metric factor $A(x)$ 
obtained here coincides (within error bars) 
with that obtained from the fit of $\xi/L$
which is consistent with the expectation from Eq.~\ref{FSS1}. 
A further check for the obtained value of $\eta$ is provided 
by the behavior of $\chi_{SG}/L^{2-\eta}$ as a function of $T$ for various $L$ at 
a fixed $x$. A value of $\eta \approx 0.22$ gives a crossing point in 
$T$ in agreement with the estimate obtained from the data of $\xi/L$ 
(see Fig.~\ref{fig6b} for the case of $x=1/128$). 
The finite size scaling procedure is summarized in Appendix~\ref{appendixD}.

\underline{\it Conclusions:} We have studied an emergent anisotropic 
dipolar system of Ising spins that arises when dipolar 
spin ice is weakly diluted with a fraction $x$ of 
non-magnetic impurities in the 
three-dimensional pyrochlore lattice. These emergent Ising spins 
have orientations that are neither random nor collinear but are 
picked according to the local easy axes of the occupied sites. This problem 
provides a lattice realization for studying the universal physics 
of a possible Ising spin glass transition in 
three dimensions, thus complementing the known cases of spin freezing for 
random dipoles, namely dense dipoles on a cubic lattice with random 
orientations~\cite{denseglass}, or dilute but collinear dipoles 
on a cubic lattice~\cite{Gingras, AndresenKOS14}.

Metropolis algorithm supplemented by parallel tempering in temperature is 
unable to equilibrate this problem except for a small number of dipoles 
because of the rapidly increasing autocorrelation time caused by rare-region 
effects of strongly interacting spin clusters. 
We use an improved cluster algorithm to relieve these equilibration 
bottlenecks and simulate much larger number of dipoles than possible using 
less elaborate algorithms.
  
Using finite-size scaling, we have been able 
to establish a finite temperature phase transition at small $x$. Furthermore, 
we present strong evidence that at small $x$, the universality class of the 
transition is independent of $x$ 
and estimate the critical exponents to be $\nu=1.27(8)$ 
and $\eta=0.228(35)$. The estimation of both the exponents $\nu$ and $\eta$ is 
a first for such dilute dipoles in three dimensions.

Our algorithm is also expected to give a significant 
speed-up for other Ising systems with atypical strong bonds, e.g., 
the recently introduced random Coulomb antiferromagnet in three 
dimensions~\cite{RehnMY16}.
Finally, beyond the thermodynamic phase transition, a detailed 
understanding of the nature of the dynamical slowdown and the resulting 
spatial heterogeneity in the local spin relaxation~\cite{BiltmoH12}
above the transition temperature $T_x$ 
for any local dynamics (Fig.~\ref{fig1a}, Fig.~\ref{fig1b}) 
remains an interesting open problem~\cite{ongoingwork}.

\underline{\it Acknowledgements:} 
We thank A.~P.~Young and Pushan Majumdar for several useful discussions.
A.S. is partly supported through the Partner Group program between the Indian 
Association for the Cultivation of Science (IACS), Kolkata and the Max Planck 
Institute for the Physics of Complex Systems (MPIPKS), Dresden. 
The simulations were performed on the supercomputing clusters 
maintained by the Max Planck Computing and Data Facility (MPCDF). 
This work was in part 
supported by Deutsche Forchungsgemeinschaft (DFG) via grant SFB 1143
as well as Cluster of Excellence ct.qmat (EXC 2147, project id 39085490).

\bibliographystyle{revtex}
\bibliography{references}

\appendix

\section{Dependence of thermalization timescale on ($a_s,b_s,C_L$)}
\label{appendixA}
The cluster construction requires specification of three parameters 
$a_s,b_s$ and $C_L$ as described in the main text. For most of the 
simulations reported in the paper, we take $a_s=1.3125$, $b_s=0.75$ and 
$C_L=N/5$ which is enough to make the last three bins agree for $\xi/L$ and 
$\chi_{SG}$ when performing logarithmic binning tests for equilibration. 
However, this is not true for the case of $L=22$ for the smallest dilution 
of $x=1/128$ that we considered (as can be seen from Fig.~\ref{fig7a}). 
Here, we instead see that choosing $a_s=1.3125$, $b_s=47.25$ and $C_L=N/3$ 
and taking the probability of a cluster flip to be $50\%$ (instead of $15\%$) 
improves the equilibration significantly (Fig.~\ref{fig7a}) and we use these 
parameters to also generate the data at $L=22,x=1/128$. 

In our cluster construction procedure, the value of $\Delta J = b_s J_s$ 
scales linearly with $x$ if $a_s$ is kept fixed. As a result, 
unless $b_s$ is increased with decreasing $x$, the number of cluster sets 
increases for highly diluted systems at very low $x$. 
This is not ideal since (a) the different cluster 
sets are then not {\it different enough} and (b) with our considered 
probability distribution, the first few sets, i.e., $C_0$ 
and its neighboring sets are hardly chosen during the cluster flip. 
Thus, it is useful to increase $b_s$ as one goes to 
lower values of $x$. Furthermore, at low $x$, $T_x$ also decreases with 
$x$ and hence the size of the largest frozen cluster (with respect to 
single spin flips) is also bound to increase. We therefore increase 
$b_s$ significantly from $0.75$ to $47.25$ and $C_L$ from 
$N/5$ to $N/3$ for $L=22, 
x=1/128$. We also see that at low $x$, it is better to increase the 
relative probability of cluster flips with respect of single spin flips and 
therefore increase this from $15\%$ to $50\%$. 
The problem of optimizing over the parameters $a_s$, $b_s$ and $C_L$ has 
not been systematically addressed in this work and understanding this 
should lead to further significant speed-up at very low $x$ 
as Fig.~\ref{fig7a} already demonstrates.

\section{Role of different cluster sets in equilibration}
\label{appendixB}
\begin{figure*}[t]
\centering
\subfigure[\label{fig7a}]{\includegraphics[scale=0.23]{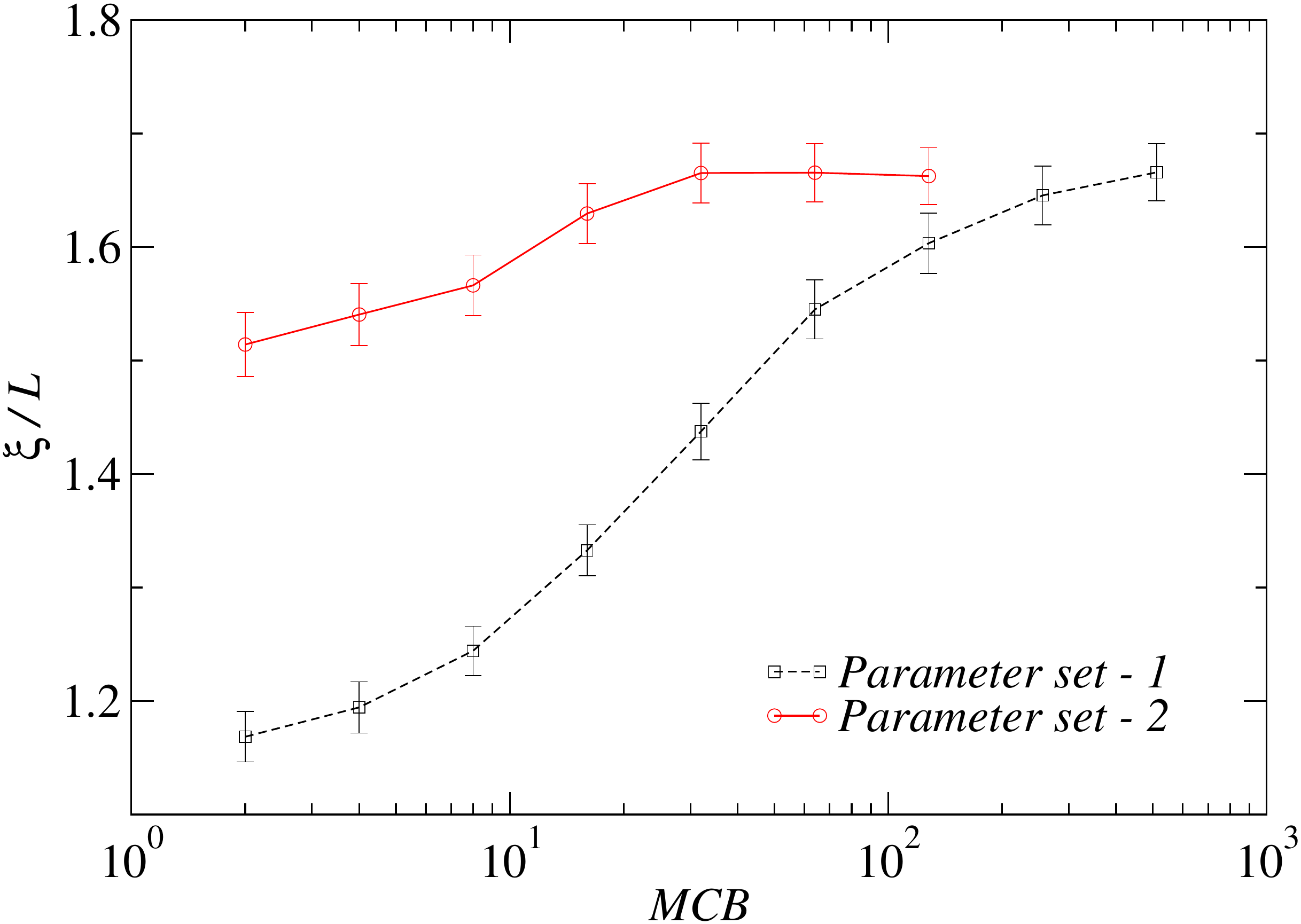}}
\subfigure[\label{fig7b}]{\includegraphics[scale=0.23]{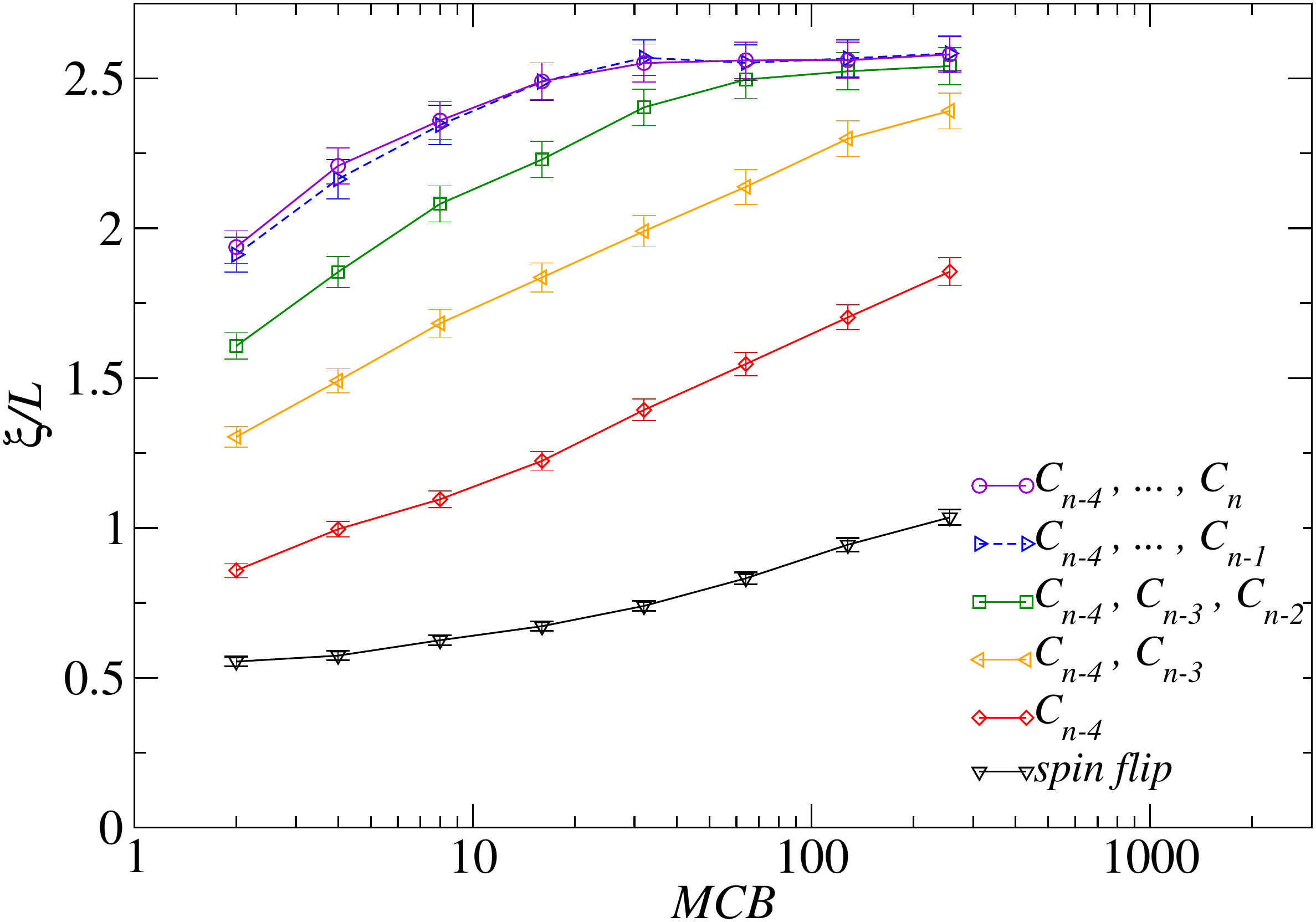}}
\subfigure[\label{fig7c}]{\includegraphics[scale=0.23]{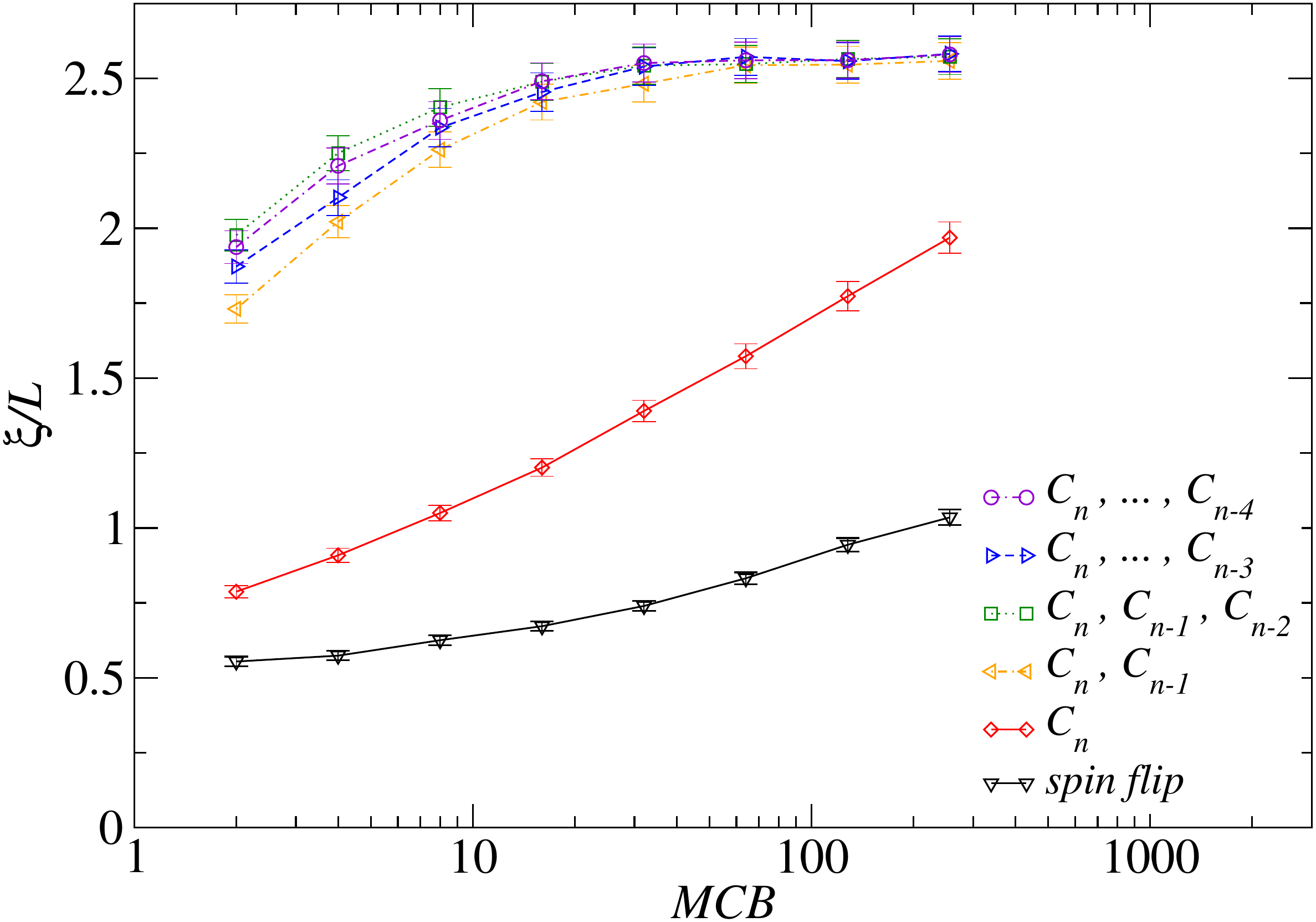}}
\subfigure[\label{fig7d}]{\includegraphics[scale=0.45]{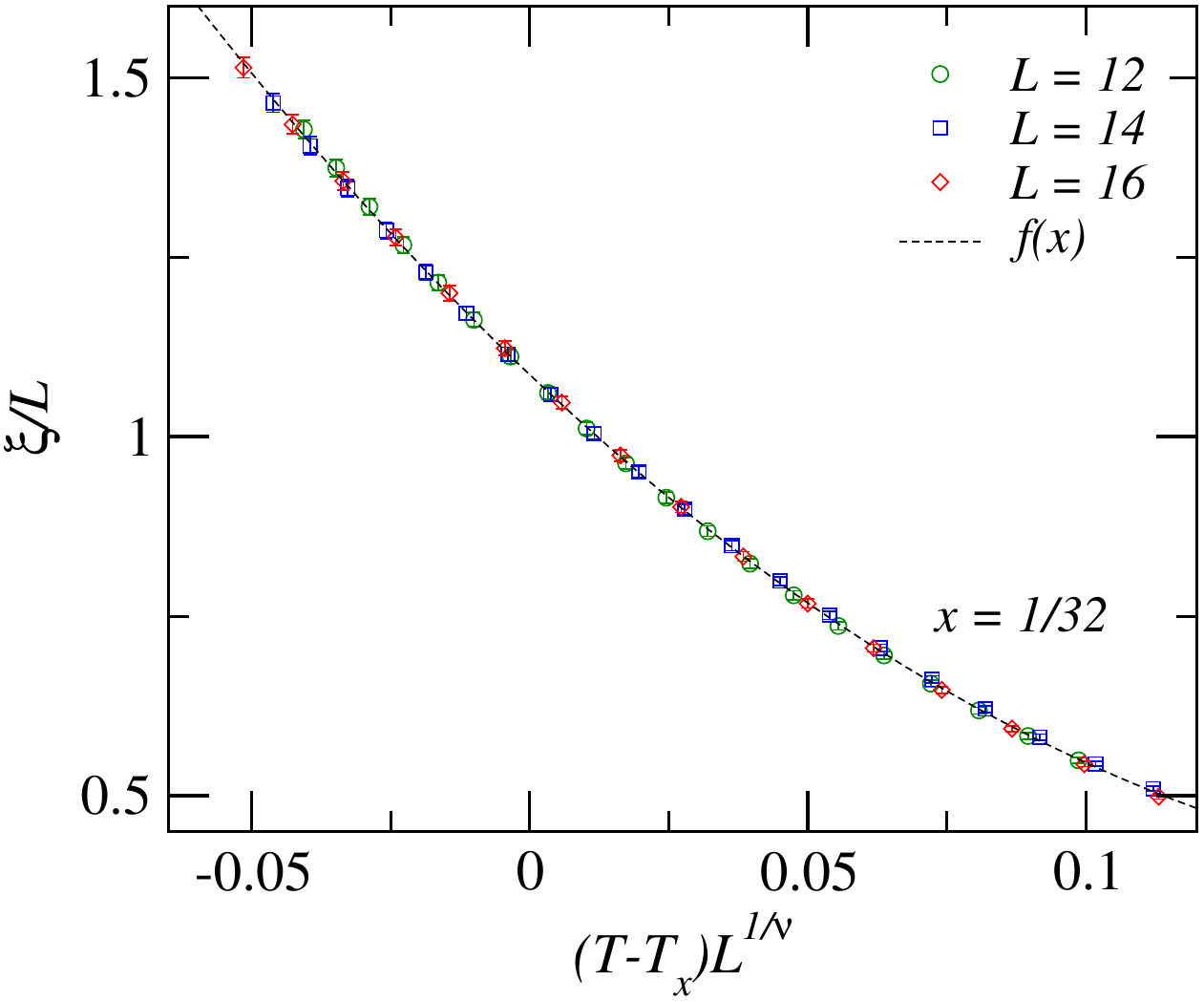}}
\subfigure[\label{fig7e}]{\includegraphics[scale=0.45]{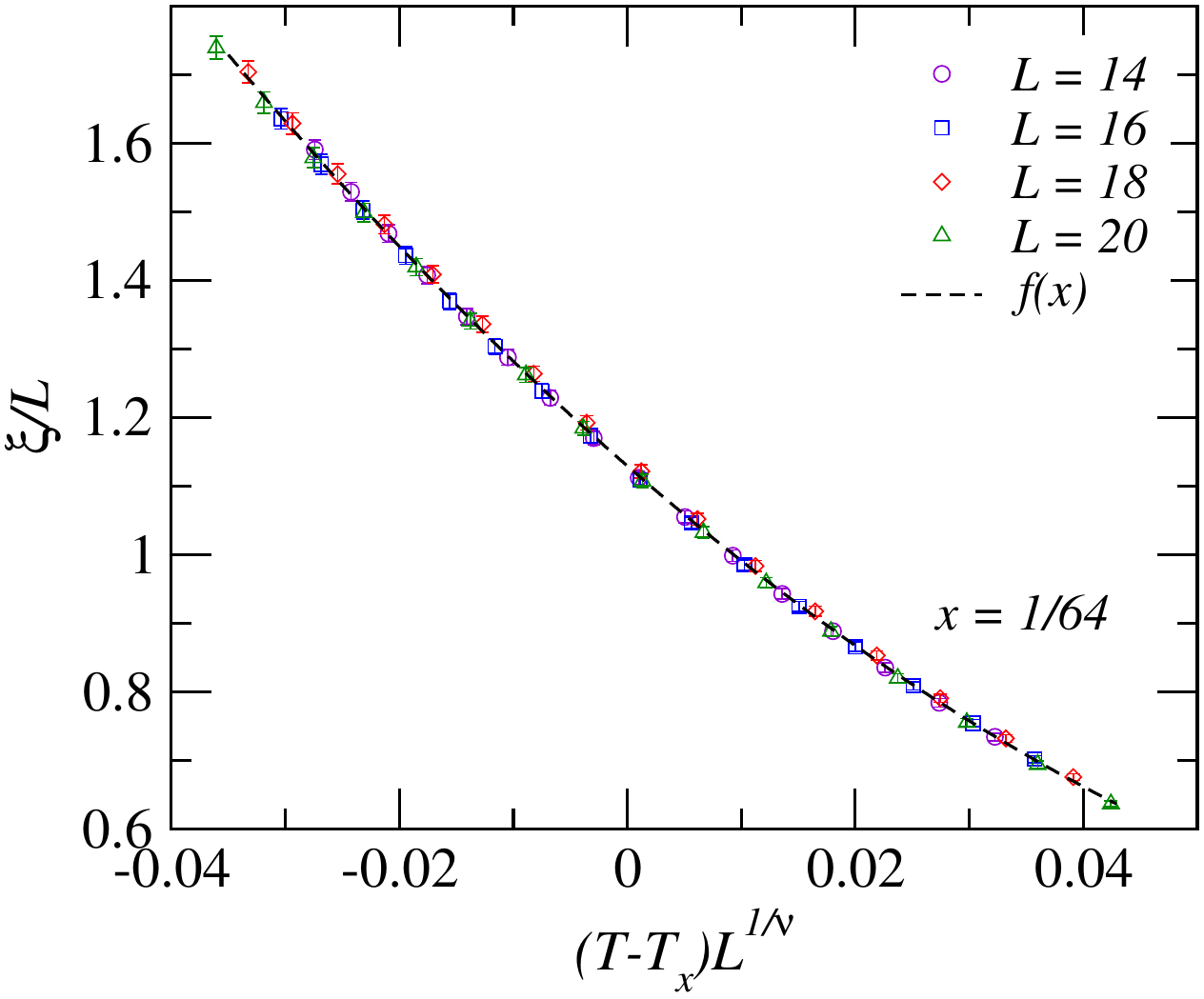}}
\subfigure[\label{fig7f}]{\includegraphics[scale=0.45]{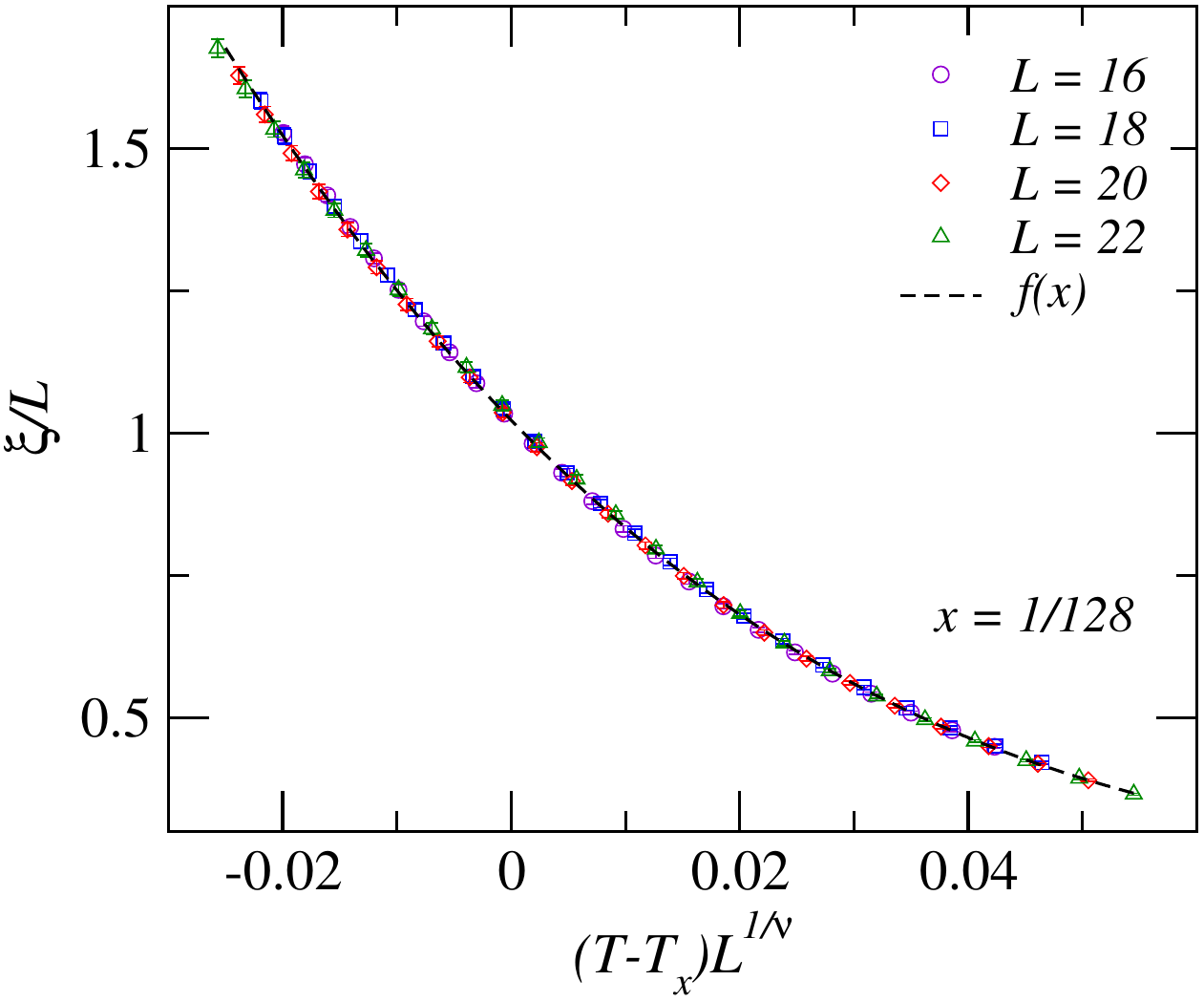}}
\caption{{\label{fig7}}{(a) Comparison of the logarithmic binning for $\xi/L$ for $L = 22$ at $x=1/128$ using parameter set-1 ($a_s=1.3125, b_s=0.75, C_L=N/5$) 
and parameter set-2 ($a_s=1.3125, b_s=47.25, C_L=N/3$) 
with the local spin flip move chosen $85\%$ ($50\%$) in the former (latter) 
case. The disorder averaging is done over $N_{sample}=600$, 
and the parallel tempering parameters are $T_0=0.00675$, $c=0.031$ and $N_T=31$ (the result is shown for the lowest temperature $T_0$). Logarithmic binning thermalization test for $\xi/L$ for $L=10$, $x=1/32$ at a low temperature of $T=0.015 (\lesssim T_{x}/2)$ 
where the cluster sets are switched on from $C_{n-4}$ to $C_n$ in (b) and 
from $C_n$ to $C_{n-4}$ in (c). The relative probabilities to pick the 
individual cluster sets is given in Table~\ref{tabI}. 
In panels (a), (b), (c), one MCB equals 512 MC Sweeps in all the figures. The data collapse of $\xi/L$ at (d)$x=1/32$ (e) $x=1/64$ and (f) $x=1/128$. The extracted values of $T_x$ and $\nu$ are summarized in Table~\ref{tabIII}.
 }
 }
\end{figure*}

%We now come to the issue of the relative probability of choosing a cluster 
%set. In the previous approach~\cite{AndresenKOS14, JanzenHE08, AndresenJK11}, 
%a cluster was uniformly picked out of all the possible cluster sets 
%$C_0,C_1,\cdots,C_n$ before the cluster move was attempted. In our approach, 
%we pick the clusters in the set $C_n$ (where the clusters contain the largest 
%fraction of the strong bonds) with the highest probability and the clusters 
%in the set $C_0$ with the lowest probability to aid faster equilibration of 
%the system. 
We explictly demonstrate the role of the different cluster 
sets in equilibrating the system by considering $N_{sample}=300$ 
independent disorder realizations for a system of size $L=10$ with $x=1/32$ 
at a rather low temperature of \(T=0.015(\lesssim T_{x}/2)\) 
(Fig.~\ref{fig7b} and Fig.~\ref{fig7c}). 
The cluster sets are constructed for each disorder realization by 
taking the parameters to be $a_s=1.3125, b_s=0.75$ and $C_L=N/3$. 
The total number of cluster sets, $(n+1)$, is $5-6$ for most of the 
disorder realizations. We then create various cluster set combinations 
(given in Table~\ref{tabI}) by either switching on the cluster sets one by one 
from $C_{n-4}$ to $C_n$ or from $C_{n}$ to $C_{n-4}$ (in the disorder 
realizations where $n>4$, the cluster sets $C_{n-5}$ and lower are not 
considered for this analysis). 
The parameters for the parallel tempering used are $T_0=0.015, c=0.065$ 
and $N_T=31$. We choose a local spin (cluster) flip 
with probability $85\%$ ($15\%$). The (relative) probability to choose a 
particular cluster set $C_m$ is then taken to be $P_m$ (see Table~\ref{tabI}) 
according to the rule specified in the main text. We check the difference in 
thermalization time for the various cluster set combinations by performing
a logarithmic binning analysis for $\xi/L$ and the results are displayed in 
Fig.~\ref{fig7b} and Fig.~\ref{fig7c}.

It can be seen that neither $C_{n-4}$, $C_{n-4}$, $C_{n-3}$
nor $C_{n-4}$, $C_{n-3}$, $C_{n-2}$ satisfy the log binning 
thermalization test within the given number of MCS but their performance 
improves progressively. However, when the sets containing larger fraction of 
strong bonds are considered in $C_{n-4}$, $\cdots$, $C_{n-1}$  
and $C_{n-4}$, $\cdots$, $C_n$, 
the system does equilibrate within the given number of MCS (Fig.~\ref{fig7b}). 
What happens when we start switching on the clusters from $C_n$ to $C_{n-4}$? 
Here the effect is much more dramatic and already the cluster 
combination of $C_n$, $C_{n-1}$ equilibrates the system for the given number of 
MCS (note that the same is not true with just the cluster set $C_n$). 
The equilibration performance only improves slightly when we consider the 
combinations $C_n$, $C_{n-1}$, $C_{n-2}$, $C_n$, $\cdots$, $C_{n-3}$ and 
$C_n$, $\cdots$, $C_{n-4}$ (Fig.~\ref{fig7c}). 
These results clearly show that it is 
important to attempt flipping clusters from the latter sets  
more frequently than to attempt flipping clusters from the earlier sets.
In this manner, the average computational cost of $1$ MCS is 
also reduced as the larger
clusters are flipped less often than the smaller clusters.

%%%%%%%%%%%%%%%%%TABLE I
\begin{table}[htbp]
  \centering
\vspace{0.5cm}
  \begin{ruledtabular}
  \begin{tabular}{c|c}
    Cluster Combination & Relative Probabilities \\
 \hline
$C_{n-4}$ & 1 \\
$C_{n-4}, C_{n-3}$ & 1/2,1/2 \\
$C_{n-4}, C_{n-3},C_{n-2} $ & 1/4,1/4,1/2 \\
$C_{n-4}, C_{n-3},C_{n-2}, C_{n-1}$ &  1/8,1/8,1/4,1/2\\
$C_{n-4}, C_{n-3},C_{n-2}, C_{n-1}, C_n$ & 1/16,1/16,1/8,1/4,1/2\\
$C_n$ & 1 \\
$C_n, C_{n-1}$ & 1/2,1/2 \\
$C_n, C_{n-1}, C_{n-2}$ & 1/2,1/4,1/4 \\
$C_n, C_{n-1}, C_{n-2}, C_{n-3}$ & 1/2,1/4,1/8,1/8 \\
$C_n, C_{n-1}, C_{n-2}, C_{n-3}, C_{n-4}$ & 1/2,1/4,1/8,1/16,1/16
\end{tabular}
\end{ruledtabular}
\caption{The relative probabilities to pick the individual cluster sets which have been used to generate the results of Fig.~\ref{fig4}(a),(b)}
\label{tabI}
\end{table}

%%%%%%%%%%%%%%%%%%%%%%%%%%%%%%%%%%%%%%%%%%%%%%%%%%%%%
%***************************************

\section{Simulation parameters}
\label{appendixC}
Apart from $a_s,b_s,C_L$, the other simulation parameters that we need to 
specify are $T_0$, $c$ and $N_T$ to set up the parallel tempering protocol. 
We summarize the values of these parameters for different $L$ and $x$ and 
also $N_{MCS}$ (the number of MCS used), 
$N_{sample}$ used during the production runs in Table~\ref{tabII}.
%%%%%%%%%%%%%%%%%TABLE II
\begin{table}[htbp]
  \centering
\vspace{0.5cm}
  \begin{ruledtabular}
  \begin{tabular}{c|c|c|c|c|c|c}
    $x$ & $L$ & $T_0$ & $c$ & $N_T$ & $N_{MCS}$ & $N_{sample}$ \\
 \hline
1/32 & 4 & 0.03 & 0.20 & 15 & $10\times2^{19}$ & 2500 \\
1/32 & 6 & 0.03 & 0.035 & 63 & $10 \times 2^{18}$ & 1500 \\
1/32 & 8 & 0.03 & 0.035 & 63 & $10 \times 2^{17}$ & 1500 \\
1/32 & 10 & 0.03 & 0.025 & 63 & $10 \times 2^{16}$ & 1500 \\
1/32 & 12 & 0.03 & 0.025 & 63 & $10 \times 2^{17}$ & 1500 \\
1/32 & 14 & 0.03 & 0.025 & 63 & $10 \times 2^{16}$ & 1500 \\
1/32 & 16 & 0.03 & 0.030 & 31 & $10 \times 2^{17}$ & 1500 \\
1/64 & 4 & 0.01 & 0.055 & 31 & $10 \times 2^{17}$ & 2100 \\
1/64 & 6 & 0.01 & 0.055 & 31 & $10 \times 2^{17}$ & 2100 \\
1/64 & 8 & 0.01 & 0.055 & 31 & $10 \times 2^{16}$ & 2100 \\
1/64 & 10 & 0.01 & 0.040 & 31 & $10 \times 2^{17}$ & 1500 \\
1/64 & 12 & 0.01 & 0.040 & 31 & $10 \times 2^{17}$ & 1500 \\
1/64 & 14 & 0.0135 & 0.031 & 31 & $10 \times 2^{17}$ & 1500 \\
1/64 & 16 & 0.0135 & 0.031 & 31 & $10 \times 2^{18}$ & 1500 \\
1/64 & 18 & 0.0135 & 0.031 & 31 & $10 \times 2^{17}$ & 1500 \\
1/64 & 20 & 0.0135 & 0.031 & 31 & $10 \times 2^{17}$ & 1500 \\
1/128 & 6 & 0.00675 & 0.045 & 31 & $10 \times 2^{17}$ & 2500 \\
1/128 & 8 & 0.00675 & 0.045 & 31 & $10 \times 2^{17}$ & 2500 \\
1/128 & 10 & 0.00675 & 0.031 & 31 & $10 \times 2^{17}$ & 2500 \\
1/128 & 12 & 0.00675 & 0.031 & 31 & $10 \times 2^{17}$ & 2500 \\
1/128 & 14 & 0.00675 & 0.031 & 31 & $10 \times 2^{16}$ & 2200 \\
1/128 & 16 & 0.00675 & 0.031 & 31 & $10 \times 2^{17}$ & 1500 \\
1/128 & 18 & 0.00675 & 0.031 & 31 & $10 \times 2^{20}$ & 1500 \\
1/128 & 20 & 0.00675 & 0.031 & 31 & $10 \times 2^{20}$ & 1500 \\
1/128 & 22 & 0.00675 & 0.031 & 31 & $10 \times 2^{18}$ & 1500 
\end{tabular}
\end{ruledtabular}
\caption{Parameters of the Monte Carlo simulations.}
\label{tabII}
\end{table}
\section{Finite-size scaling}
\label{appendixD}
To estimate the $T_x$ and extract the critical exponents $\nu$ 
and $\eta$, we use the scaling forms given in Eq.~\ref{FSS1} sufficiently 
close to the critical point. For $\xi/L$, we expand the scaling function 
$\mathcal{F}_1(X)$ (where $X=A(x)(T-T_x)L^{1/\nu}$) as a third-order 
polynomial $\mathcal{F}_1(X)\equiv f(X)=a_0+a_1X+a_2X^2+a_3X^3$ and 
then perform a global fit to determine the unknown parameters 
$(a_0,a_1,a_2,a_3,a,b,\nu,A(x=1/64),A(x=1/128))$ 
[where $T_x=ax(1+bx)$ is assumed at small $x$ and the metric
 factor $A(x=1/32)=1$] by minimising the reduced chi square per degree of 
freedom $\bar{\chi}^2$ defined by 
\begin{eqnarray}
\bar{\chi}^2 = \frac{1}{N_d-M} \sum_{i=1}^{N_d} (y_i -f(X_i))^2/\sigma_i^2
\label{chisq}
\end{eqnarray}
where $N_d$ equals the total number of data points, $M$ denotes the 
number of fitting parameters, $y_i$ denotes the mean value of the $i$-th 
data point, $\sigma_i$ denotes the error in the $i$-th data point and 
$f(X_i)$ denotes the fitting function. The fits are considered of good quality 
when $\bar{\chi}^2 \lessapprox 1$. Since all temperatures are simulated 
with the same disorder realization in the parallel tempering procedure, the 
fitted data is correlated. We therefore apply a bootstrap analysis to the data 
to estimate the statistical error bars on the various fit parameters. 
It is useful to emphasize here that the quoted error bars are only statistical 
errors since estimating systematic errors properly requires a reliable 
knowledge of the corrections due to scaling. For $\chi_{SG}$, we again use 
the scaling form given in Eq.~\ref{FSS1}(b) and expand the scaling function 
$\mathcal{F}_2$ 
as a third-order polynomial. We further fix the values of $a,b$ that determine 
$T_x=ax(1+bx)$ and the exponent $\nu$ from the previous fit of $\xi/L$ and 
then perform the minimization of $\bar{\chi}^2$ to determine the exponent 
$\eta$ and the other fitting parameters.

%%%%%%%%%%%%%%%%%%%%%%%%%%%%%%%%%%%%%%%%%%%
We show the data collapse of $\xi/L$ at each individual $x$ in 
Fig.~\ref{fig7d}, Fig.~\ref{fig7e} and Fig.~\ref{fig7f} for 
completeness. The extracted values of $T_x$ and $\nu$ are shown in 
Table~\ref{tabIII} and 
is fully consistent with the scenario that $T_x \propto x$  
and the universality of the critical point is independent of $x$ for small 
$x$.  
%%%%%%%%%%%%%%%%%TABLE III
\begin{table}[htbp]
  \centering
\vspace{0.5cm}
  \begin{ruledtabular}
  \begin{tabular}{c|c|c|c}
    $x$ & $T_x$ & $\nu$ & $\bar{\chi}^2$ \\
 \hline
1/32 & 0.0351(8) & 1.21(6) & 0.73 \\
1/64 & 0.0171(4) & 1.30(7) & 0.98 \\
1/128 & 0.0090(2) & 1.26(4) & 1.06
\end{tabular}
\end{ruledtabular}
\caption{$T_x$, $\nu$ and $\bar{\chi}^2$ obtained from the data collapse of $\xi/L$ at $x=1/32$, $x=1/64$ and $x=1/128$ (see Fig.~\ref{fig7}(d),(e),(f)).}
\label{tabIII}
\end{table}
%%%%%%%%%%%%%%%%%%%%%%%%%%%%%%%%%%%%%%%%%%%%%%%
\end{document}